\documentclass[prc,twocolumn,showpacs,preprintnumbers,amsmath,amssymb]{revtex4-1}
\usepackage[utf8]{inputenc}
\usepackage{graphicx}
\usepackage{tabularx}
\usepackage{bm}
\usepackage{color}
\def\mod#1{{\bf\color[rgb]{1,0,0} #1}}

\usepackage[normalem]{ulem}  
\renewcommand\sout{\bgroup \color{red} \ULdepth=-.5ex \ULset}
\usepackage{hyperref}

\allowdisplaybreaks

\begin{document}

\title{Model selection for $K^+\Sigma^-$ photoproduction within isobar model}

%

\author{P.~Byd\v{z}ovsk\'{y}$^{1}$, A.~Ciepl\'{y}$^{1}$, D.~Petrellis$^{1}$, D.~Skoupil$^{1}$, N.~Zachariou$^{2}$}

\affiliation{$^{1}$Nuclear Physics Institute, CAS, \v{R}e\v{z}/Prague, 
Czech Republic\\
$^{2}$University of York, York YO10 5DD, United Kingdom
}

\date{\today }

\begin{abstract}

We utilise an isobar model to investigate the $K^+ \Sigma^-$ photoproduction off a neutron in the resonance region. Except for the Born terms, we include high-spin (spin-3/2 and spin-5/2) nucleon resonances in the consistent formalism together with a few $\Delta$ and kaon resonances to achieve an acceptable agreement with data. Interestingly, we reveal that no hyperon resonances are needed to achieve a reasonable description of data. On the other hand the $N(1720)3/2^+$ resonance was found to be very important for correct description of data. The free parameters of the model were fitted to experimental data from the LEPS and CLAS Collaborations on either differential cross sections or photon beam asymmetry. The novel feature of the fitting procedure is the use of a regularization method, the Least Absolute Shrinkage Selection Operator, and information criteria in order to choose the best fit.

\end{abstract}


\maketitle
\section{Introduction}
\label{sec:introduction}

The study of the kaon-hyperon photo and electroproduction from nucleons in the third nucleon resonance region provides important information about the spectrum of baryon resonances and interactions in the systems of hyperons and nucleons, which arise from Quantum Chromodynamics. Not only do we aim at studying the reaction mechanism, we also focus on obtaining more information about the existence and properties of the so-called ``missing" resonances, which have been predicted by quark models but have not been seen in the pion production of $\pi N$ scattering processes~\cite{CapRob00,Bonn}. These states may have escaped experimental confirmation due to their stronger decay coupling to $K\Lambda$ and $K\Sigma$ rather than to the more well-known pion final states, see results of the coupled-channel analysis~\cite{Julich-Bonn} and outcomes of the partial-wave analysis~\cite{Anisovich}.

A plethora of theoretical studies on hyperon production have been performed over the  past decades 
with focus primarily given to the $K^+\Lambda$ production channel off the proton due to the large amount of available experimental data, see e.g.  Ref.~\cite{Skoupil:2016ast} and references therein. The analyses before 2004 unfortunately suffered from a lack of high-quality experimental data~\cite{AS90} but the situation changed dramatically after new high-duty-factor accelerators, providing good quality, high-current, and polarized continuous beams, were constructed in Jefferson Lab (CEBAF) and Bonn University (ELSA). 

For the time being, the database of the channels using neutron targets is very limited, with available measurements of the differential cross section for $K^+\Sigma^-$~\cite{LEPS-Kohri, CLAS-Pereira} and $K^0\Lambda$ reactions~\cite{CLAS-Pereira}. Inclusive momentum spectra in $K^0$ photoproduction off deuteron were measured in the threshold region at LNS of Tohoku University~\cite{Tsu08}. There are only two measurements of the beam asymmetry $\Sigma$: from LEPS~\cite{LEPS-Kohri} 
associated with very limited kinematical coverage, and just recently a precise measurement from the CLAS collaboration~\cite{CLAS21}, which covers a wide range of kinematics. 
There are also recent results on beam-target helicity asymmetry $E$, also from the CLAS collaboration~\cite{CLAS-E}. 

In the present paper we analyse the new CLAS data on the asymmetry 
together with the other older data in the $K^+\Sigma^-$ channel using 
an isobar model, similarly as we did it in Ref.~\cite{Skoupil:2016ast} in 
the $K^+\Lambda$ channel, to extract information on the resonance content 
of the model and its parameters. 
However, in the present analysis, we elaborate on the method of adjusting 
free parameters of the model which allows a better description of the available experimental data and we also deem that it is more sensitive to a selected 
resonant content of the model. 
One version of our model, Fit M, see the next Sect.,  
was also utilised in Ref.~\cite{CLAS21} for comparison with the data. 

The paper is organized as follows: In Section II., we discuss the isobar 
model which we use for describing the 
$\Sigma^-$ photoproduction reaction off the neutron. Section III. deals with the free parameters in the model
and with the new method of their fitting to the data.
In Section IV. we discuss the obtained results 
and Section V. 
provides a brief summary and conclusions.

\section{Model description}
\label{sec:model}

The current model based on an effective Lagrangian in the tree-level 
approximation is constructed to describe data only in the $K^{+}\Sigma^{-}$ channel
assuming no final-state interaction. 

The non resonant part of the amplitude consists of the Born terms and
exchanges of resonances in the $t$  channel (K* and K$_1$ kaon resonances) and $u$ channel ($\Sigma^{*}$ hyperon resonances).
The main coupling constant $g_{K^+\Sigma^- n}=\sqrt{2}\; g_{K^+\Sigma^0 p}$, that determines the strength of the Born
terms, was taken from the model constructed for  the $K^{+}\Lambda$ channel \cite{Skoupil:2016ast} and kept unchanged in the
present fit. The resonant part is modelled by s-channel exchanges of nucleon
and $\Delta$ resonances with masses from around the process threshold up to about 2 GeV. 
Hadronic form factors included in the strong vertexes account for a hadron
structure and regularize the amplitude at large energies. The form factors
are introduced in the way that keeps gauge invariance in analogy with the
method used in Refs.~\cite{Skoupil:2016ast} and \cite{Skoupil:2018vdh}. The relevant formulas to show gauge invariance of the  amplitude in the case of $K^+\Sigma^-$ photoproduction are given in Appendix A.

The considered set of nucleon resonances was motivated by previous analyses of $K^{+}\Lambda$ 
and $K\Sigma$ photoproduction \cite{Skoupil:2016ast, Skoupil:2018vdh} and \cite{David:1995pi}, 
respectively. Some additional nucleon resonances decaying strongly into the $K\Sigma$ channel 
were also considered in our analysis, together with $\Delta$ and $\Sigma$ resonances used to complement 
the model in the $s$-channel and $u$-channel sectors. The free parameters of a particular model, 
the coupling constants, and ranges of hadronic form factors for a given set of resonances, were fitted 
to the data, and the quality of the model was checked by comparing its prediction with the data. 
In the end, a variant with the smallest $\chi^{2}/\text{n.d.f.}$ and reasonable values of the parameters 
was selected. 
In our current best fit with the CERN Minuit library~\cite{MINUIT}, which is also presented in Ref.~\cite{CLAS21} and which was aimed at description of the new CLAS data on asymmetry, we have used 14 resonances which are shown in Table \ref{tab:BCSfit}. It includes the *** and **** baryon states with most of them decaying into the $K\Lambda$ and $K\Sigma$ channels \cite{PDG:2020xxx} and also some other $\Delta$ and $\Sigma$ resonances which we have taken into account in our analysis. We note that we used only the statistical errors when computing the $\chi^2$, which results in a relatively large value $\chi^{2}/\text{n.d.f.} = 2.39$ obtained for the selected solution. The reason for using only statistical errors was missing systematic errors in some data sets. When systematics is taken into account, the $\chi^2$ value usually drops, but the inclusion of systematics does not change the quality of results. 

In the following section we will introduce the fitting procedure and its extension --  the Least Absolute Shrinkage Selection Operator (LASSO) -- which is a more sophisticated method for adjusting the free model parameters. The LASSO method was used in a recent analysis of pion photoproduction and study of baryon resonances~\cite{Landay}. The advantage of this method lies in its capability of removing redundant parameters and thus limiting the number of contributing resonances.

\begin{widetext}

\begin{table}[h]
\caption{Characteristics of included resonances with their masses and widths taken 
as the PDG Breit-Wigner averages. The available branching ratios to the $K \Lambda$ 
and $K \Sigma$ channels are also taken from the PDG \cite{PDG:2020xxx}. For the nucleon and $\Delta$ resonances, the values of coupling constants, $g_1$ and $g_2$, show the baryon-$K\Sigma$ scalar and tensor couplings obtained in our fit, while for the $K^{*}$ and $K_1$ states they represent the vector and tensor couplings, respectively. We show values of $g_1$ and $g_2$ achieved with the Minuit only (denoted as fit M) and with the LASSO method (fit L).}
\begin{center}
\begin{tabularx}{\textwidth} {llllXXXXXX}
\hline \hline
Nickname & Resonance      &  Mass      &  Width  & \multicolumn{2}{c}{Branching ratios}
                                      & \multicolumn{2}{c}{Fit M cc's}
                                      & \multicolumn{2}{c}{Fit L cc's} \\
     &               & (MeV)    & (MeV)   & $K \Lambda$ & $K \Sigma$ &  $g_1$  &  $g_2$  &  $g_1$  &  $g_2$  \\
\hline
K*& $K^*(892)$       & 891.7    & 50.8    &     ---     &    ---     &  0.366  &  1.103  &  0.3095 &  ---    \\ 
K1& $K_1(1270)$      & 1270     & 90      &     ---     &    ---     & $-1.448$&  0.473  &  ---    &  ---    \\ 
N3& $N(1535)\;1/2^-$ & 1530     & 150     &     ---     &    ---     & $-0.709$&   ---   &  ---    &  ---    \\
N4& $N(1650)\;1/2^-$ & 1650     & 125     &     0.07    &    0.00    &  0.314  &   ---   &$-0.0848$&  ---    \\
N8& $N(1675)\;5/2^-$ & 1675     & 145     &     ---     &    ---     & $-0.013$&  0.022  &$-0.0094$&  0.0033 \\
N6& $N(1710)\;1/2^+$ & 1710     & 140     &     0.15    &    0.01    & $-0.940$&   ---   &  ---    &  ---    \\
N7& $N(1720)\;3/2^+$ & 1720     & 250     &     0.05    &    0.00    & $-0.098$&$-0.082$ &$-0.1866$&$-0.1264$\\
P4& $N(1875)\;3/2^-$ & 1875     & 200     &     0.01    &    0.01    & $-0.220$&$-0.223$ &$-0.0421$&  0.0250 \\
P1& $N(1880)\;1/2^+$ & 1880     & 300     &     0.16    &    0.14    & $-0.050$&   ---   &  ---    &  ---    \\
Mx& $N(1895)\;1/2^-$ & 1895     & 120     &     0.18    &    0.13    & $-0.063$&   ---   &  0.0192 &  ---    \\
P2& $N(1900)\;3/2^+$ & 1920     & 200     &     0.11    &    0.05    & $-0.051$&$-0.004$ &  0.0296 &  0.0104 \\
M4& $N(2060)\;5/2^-$ & 2100     & 400     &     0.01    &    0.03    &$-0.00001$& 0.003  &$-0.0030$&  0.0042 \\
M1& $N(2120)\;3/2^-$ & 2120     & 300     &     ---     &    ---     & $-0.034$&$-0.010$  &  0.0003 &  0.0001 \\
D1& $\Delta(1900)\;1/2^-$ & 1860 & 250    &     ---     &    0.01    &  0.298  &   ---   &  ---    &  ---    \\
D2& $\Delta(1930)\;5/2^-$ & 1880 & 300    &     ---     &    ---     &   ---   &   ---   &  ---    &  ---    \\
D3& $\Delta(1920)\;3/2^+$ & 1900 & 300    &     ---     &    ---     &   ---   &   ---   &  ---    &  ---    \\
D4& $\Delta(1940)\;5/2^-$ & 1950 & 400    &     ---     &    ---     &   ---   &   ---   &  ---    &  ---    \\
S1& $\Sigma(1660)\;1/2^+$ & 1660 & 100    &     ---     &    ---     &   ---   &   ---   &  ---    &  ---    \\
S2& $\Sigma(1750)\;1/2^-$ & 1750 &  90    &     ---     &    ---     &   ---   &   ---   &  ---    &  ---    \\
S3& $\Sigma(1670)\;3/2^-$ & 1670 &  60    &     ---     &    ---     &   ---   &   ---   &  ---    &  ---    \\
S4& $\Sigma(2010)\;3/2^-$ & 1940 & 220    &     ---     &    ---     &   ---   &   ---   &  ---    &  ---    \\
\hline \hline
\end{tabularx}

\end{center}
\label{tab:BCSfit}
\end{table}

\end{widetext}

\section{Adjusting model parameters}
\label{sec:adjusing}

The model used in our study of strangeness production is an effective model, with coupling constants and cutoff values of hadron form factors not determined. Because of this, experimental data play a crucial role in fixing these parameters that enhance the predictive power of our model.

The free parameters to be adjusted are the main coupling constant, $g_{K^+\Sigma^- n}$, cutoff parameters for the hadron form factors of background and resonant terms, and the couplings of resonances introduced. Please note that the $g_{K^+\Sigma^- n}$ coupling was kept unchanged during the fitting procedure with Minuit but it was allowed to vary within the boundaries shown in Eq.~(\ref{eq:gks0n}) during the fitting process with the LASSO method. There is one free parameter for spin-1/2 resonance and two free parameters for spin-3/2 and 5/2 (nucleon and hyperon) resonances, while each kaon resonance introduces two additional free parameters (vector and tensor couplings).

We calculate the $\chi^2$, in order to check whether a given hypothesis describes the given data well. The optimum set of free parameters $(c_1,\ldots,c_n)$ for a given set of data points $(d_1,\ldots,d_N)$ is obtained by minimising the $\chi^2$, calculated as follows:
\begin{equation}
\chi^2 = \sum_{i=1}^{N}\frac{[d_i-p_i(c_1,\ldots,c_n)]^2}{(\sigma_{d_i}^{stat})^2},
\label{eq:chi^2}
\end{equation}
where $N$ is the number of data points, $n$ is the number of free parameters, and $p_i$ represents the theoretical prediction of observables (differential cross sections and photon beam asymmetry in this case) for the measured data point $d_i$.


The minimization was done with the help of least-squares fitting method making use of the Minuit library~\cite{MINUIT}. The $g_{K^+\Sigma^- n}$ coupling constant was kept inside the limit of the $20\%$ broken SU(3) symmetry~\cite{deSwart},
\begin{equation}
\sqrt{2}\cdot 0.8\leq\frac{g_{K^+\Sigma^- n}}{\sqrt{4\pi}}\leq \sqrt{2}\cdot 1.3.
\label{eq:gks0n}
\end{equation}
Moreover, the cutoff parameters of the hadron form factors for both background and resonant terms were kept inside the limits from 0.5 to 3.0~GeV, in order to avoid too soft or too hard form factors.

One of the problems that arise when fitting a theoretical model to experimental data is that of over-fitting the data. This means that\mod{,} although a more complex model (one with more parameters) may improve the fitting to the existing data, that model may fail to generalize to new data, resulting thus in a poor description of reality. The set of techniques that have been developed to combat this problem is known in the Machine Learning literature with the name `regularization'~\cite{Bishop,ESL}. Typically, regularization involves the addition of a penalty term in the error function that prevents the parameters of the model from taking large values, when the total error function is minimized. The penalty term may take various forms and the amount by which it contributes to the total error function is determined by the coefficient multiplying it, called regularization parameter and commonly denoted by $\lambda$. Higher values of $\lambda$ tend to push more parameters close to zero, or even exactly to zero, thus favouring simpler models (with fewer parameters), which may under-fit the data. With different values of $\lambda$ leading to different sets of parameters, and hence different models, the choice of the optimal $\lambda$ becomes a problem of model selection. For this choice we intend to use criteria based on information theory, like the Akaike and the Bayesian information criteria~\cite{ICSM}, that have been recently used in similar problems~\cite{Landay}. 

Generally speaking, the $\chi^2$ is a good measure to determine underfitting but it says nothing about overfitting~\cite{Landay}. For this reason we turn to the LASSO method in order to select the simplest model that can describe the data with the minimal amount of resonances. In order to do so, we introduce a penalty term
\begin{equation}
P(\lambda) = \lambda^4 \sum_{i=1}^{N_{res}}|g_i|,
\label{eq:penalty}
\end{equation}
where $\lambda$ is the regularization parameter, which we determine using the information criteria, $g_i$ represents couplings of resonances, and $N_{res}$ is the number of assumed resonances. We opt for the fourth power of $\lambda$\mod{,} as a higher power enables us to move quickly through the region of large values of $\lambda$ and give more weight to the region of small $\lambda$. The power affects also the step in $\lambda$, resulting in a finer sampling of the region of small $\lambda$. The reason why we want to stress the region with small $\lambda$ is that in this region more and more resonances are allowed to contribute and the results can change abruptly with only slight changes in $\lambda$. Moreover, in formula~(\ref{eq:penalty}) each resonance is penalized through its coupling $g_i$, on top of the standard definition of the $\chi^2$ in Eq.~(\ref{eq:chi^2}). In order to incorporate the penalty term, we define the penalized $\chi^2_T$ as
\begin{equation}
\chi^2_T = \chi^2 + P(\lambda).
\label{eq:chi^2_T}
\end{equation}

In practical calculations, we scan a range of $\lambda$ values and in each step minimize the $\chi^2_T$. With help of the $\chi^2_T$ values we then turn to several information criteria, which serve as a tool to determine the optimal $\lambda$ value and select the most suitable model for the description of the given data. The three information criteria used in this work are the Akaike Information Criterion (AIC)~\cite{AIC}, a finite sample size corrected version of the AIC (further on referred to as AICc)~\cite{AICc}, and the Bayesian Information Criterion (BIC)~\cite{BIC} which are respectively defined as
\begin{subequations}
\begin{align}
\text{AIC} &= 2n + \chi^2_T,\\
\text{AICc}&= \text{AIC} + \frac{2n(n+1)}{N-n-1},\\
\text{BIC} &= n\, \text{log}(N)+\chi^2_T.
\end{align}
\end{subequations}
where $n$ is the number of parameters which changes as a function of $\lambda$ and $N$ is the number of data points.

Even though it may be self-evident, let us stress that the information criteria are useful in selecting the best model in the particular set and that models which are not included in the set remain out of consideration. If all the models of the set are poor, the information criteria will still guide us towards the best model, but even that relatively best model might be poor in the absolute sense. Therefore, while using the information criteria, every effort must be made to ensure that the set of models is well founded. For detailed information on the regularization method and a brief derivation of the information criteria, see Appendix~\ref{sec:app-reg}.



\subsection{Experimental data}
In the fitting procedure, we used altogether 674 data points to fit around 20 free parameters of our model. The currently available experimental data in the $K^+\Sigma^-$ channel are data on the differential cross section, photon beam asymmetry and beam-target asymmetry only. In the last two decades, data on the differential cross section and the beam-spin asymmetry were determined by the CLAS\cite{Pereira:2009zw,CLAS21} and LEPS~\cite{Kohri:2006yx} collaborations. The CLAS collaboration provided measurements for a wide range of kinematics. The differential cross sections were measured for photon lab energies from the near-threshold value $E_\gamma^{lab}=1.15~\text{GeV}$ to $3.55~\text{GeV}$ and for $\cos \theta_K^{c.m.}$ in the range from $-0.85$ to 0.85, whereas the beam-asymmetry data span the region of photon lab energies $E_\gamma^{lab}$ from $1.1345~\text{GeV}$ to $2.276~\text{GeV}$ and $\cos \theta_K^{c.m.}$ from $-0.7687$ to $0.7484$. The LEPS collaboration provided complementary measurements at forward angles as they focused only on the region of $\cos \theta_K^{c.m.}$ from $0.65$ to $0.95$. The most recent results on the beam-spin asymmetry from the CLAS collaboration provided tight constraints due to their precision and kinematical coverage~\cite{CLAS21}. We have exploited these data, except for the beam-target asymmetry data, in the similar fashion to what we have done in Refs.~\cite{Skoupil:2016ast} and \cite{Skoupil:2018vdh} for the $K^{+}\Lambda$ channel. 

\subsection{The course of the fitting procedure}
The goal of the fitting process is to find the global minimum, i.e. the set of parameters which describe the data in the best way and produce the smallest $\chi^2$. Unfortunately, this is not an easy task as we work in a very large parameter space with numerous local minima scattered around. Thus, the results of the fitting process depend also on the starting values of the parameters that are being adjusted.

What makes the situation even worse is the fact that the $\chi^2$ is only a mathematical tool that illustrates the goodness of fit. Hence, the results with similar (or even identical) $\chi^2$ values can give rather different predictions of the observables as we may end up in different local minima. In order to distinguish satisfactory results from the unreliable ones, we pay attention not only to the final $\chi^2$ value but also to the values of fitted parameters. Moreover, we briefly check the agreement of the fit with data.

Extremely helpful in recognizing valuable outcomes of the fitting procedure is the LASSO method as described above. The first step in using this technique is initializing the resonance parameters with random values in the range from $-1$ to $+1$. The main coupling constant, $g_{K^+\Sigma^- n}$, was initialized with a random value within the range shown in Eq.~(\ref{eq:gks0n}) and the initial values for the cutoff parameters were chosen inside the range of (0.8,3.0)~GeV. We use the Forward LASSO technique which means that we decrease the value of $\lambda$ in the penalty term (see Eq.~(\ref{eq:penalty})). The starting value of $\lambda$ is chosen to be 3 and we decrease it by subsequent steps of either 0.2 or 0.1 until we reach zero. The role of this parameter, and the penalty function as a whole, rests in turning off model parameters. The larger the $\lambda$ the more model parameters are turned off. This in turn means that with large $\lambda$ we can produce very economical models with a very few parameters but their agreement with data (as illustrated by the $\chi^2$ value) is rather clumsy. This, in our opinion, is a clear sign of underfitting the data as the model includes too few parameters which do not allow it to capture the data. Usually, we arrive at reasonable fits when $\lambda$ decreases to around one. For smaller values of $\lambda$, the values of the information criteria (Eqs.(5)) tend to increase which is an indication of overfitting, i.e. introducing more parameters than is needed. In some sense, the value of $\lambda$ for the minimal value of the information criterion shows how far we are from the ideal number of parameters, i.e. $\lambda \approx 0$ would  mean that we are close to the ideal number of parameters but when $\lambda$ is significantly larger than zero, we have considered too many parameters in the beginning of the process.

What is more, there seems to be a strong influence of the $\lambda$ value on the fit results and the convergence of the fitting process. It seems that large values of $\lambda$ prevent the minimizer from converging as they turn off too many parameters. In the Minuit library, there are several different minimizers at disposal. We decided to use the Minimize minimizer as it combines the merits of Migrad and Simplex minimizers. When we include also the hyperon resonances, we observe that for $\lambda=3$ there is no convergence by either Migrad or Simplex minimizers, for $\lambda$ approximately between $0.6-2.5$ only the Simplex method converges, below 0.5 Simplex and Migrad alternate and only for $\lambda < 0.05$ the Migrad converges without Simplex being called.

In the LASSO method, we have to consider also the number of decimal digits for each parameter, i.e. from which value the parameter is considered numerically zero and therefore does not appear in the total number of fitted parameters. For example, when we do calculations with individual resonances whose coupling parameters are at the order of $10^{-5}$ or $10^{-6}$, their contributions are almost zero. Thus we reckon that we should not take so many decimal places into account. From where we stand, it seems that taking into account four decimal places for each resonance is enough. When we use more, we artificially turn on parameters that are not needed and include resonances which do not contribute. In this way we would artificially increase the number of parameters $n$ and decrease the precision of the information criteria.

The number of parameters $n$ encompasses all of the parameters we introduce, including the main coupling constant $g_{K \Sigma N}$ and the cutoffs for the background, $\Lambda_{bgr}$, and resonant terms $\Lambda_{res}$. The values of these three parameters are not included in the penalty term. The reason for doing so is that we do not want these parameters to vanish. On the contrary, the rest of the fitted parameters do appear in the penalty term which then pushes their values to zero and tends to turn off their contributions. 

Similarly to what we have done in our analysis of the $K^+\Lambda$ channel, we tried to modify the masses and widths of the nucleon resonances within the ranges provided by the PDG Tables. Unfortunately, this does not lead to any better agreement with data (even though in some cases it can produce smaller values of the $\chi^2$).

As mentioned above in the case of using pure Minuit, even when we exploit the LASSO method we may end up in one of many local minima. From what we observe, we can conclude that even with the LASSO method the minimum which we find strongly depends on the initial values of the fitted parameters. Therefore, we cannot avoid the pitfall of local minima and we cannot guarantee that the fit we end up with is the best fit which exists and can be found (which would correspond to a truly global minimum). We rather say that the models we show are among the best fits we could find.

\begin{figure}
    \centering
    \includegraphics[width=0.5\textwidth]{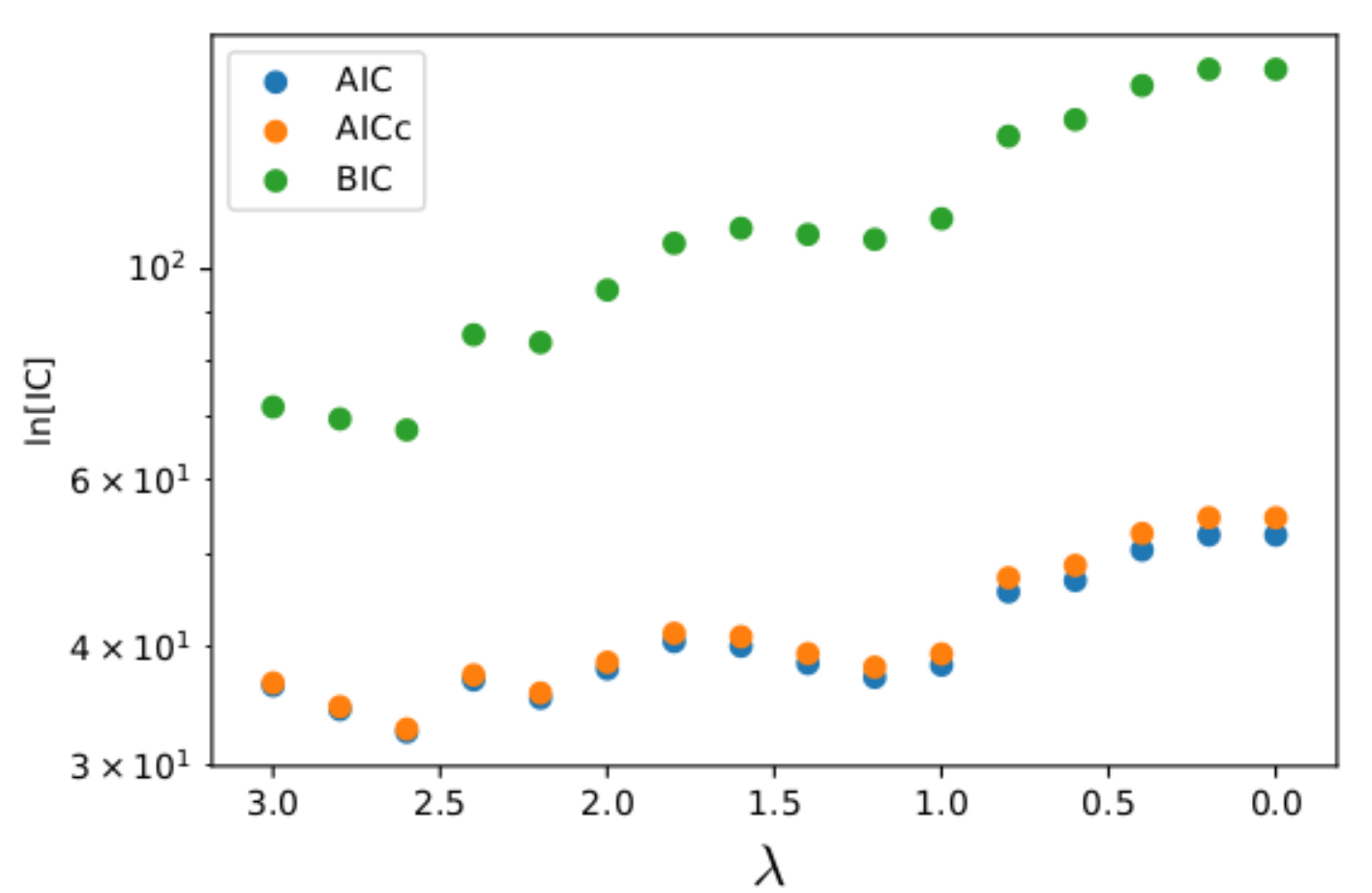}
    \caption{Information criteria values (IC) in dependence on the parameter $\lambda$. We show the Akaike (blue), corrected Akaike (orange), and Bayesian (green) information criteria. Please note that the scale of the vertical axis is logarithmic. Also note that it is not the absolute size of the individual information criteria which is important, it is the differences between IC values for different $\lambda$.}
    \label{fig:IC-sheet1}
\end{figure}

The initial set of resonances for the LASSO method was the one that was first found with help of the Minuit. We took those resonances and ran the Forward LASSO, i.e. we introduced the penalty term to the $\chi^2$ definition, as can be seen in Eq.~(\ref{eq:chi^2_T}), set the $\lambda$ at 3.0 and reduced it in subsequent steps of 0.2 until we reached $\lambda = 0$. When we plot values of the information criteria in dependence on the $\lambda$, see Fig.~\ref{fig:IC-sheet1}, we observe two distinct minima. The one is around $\lambda = 2.5$ and the other somewhere around $\lambda = 1$. We were discouraged from taking the first minimum, the deeper one, as the best fit since its $\chi^2$ is larger than the $\chi^2$ of the other minimum and the correspondence with data is also much worse. Clearly, the first minimum is a result of underfitting. The other minimum was, on the other hand, acceptable as its $\chi^2$ value appeared to be reasonable, $\chi^2/\text{n.d.f}=3.2$, and its agreement with data is good.

In subsequent fits, we added hyperon and $\Delta$ resonances on top of this core set. It seems though that no hyperon resonances are needed for data description in the $K^+\Sigma^-$ channel. None of them was found to be substantial in either the pure $\chi^2$ fitting or using the LASSO method since the couplings of hyperon resonances were zero for every value of $\lambda$. What is more, with hyperon resonances included and values of $\lambda > 0.05$ the Minuit was unable to reach convergence.

We did not rely only on the LASSO outcomes but we also compared several results with experimental data. All of the fits with the hyperon resonances tend to diverge at very forward angles. We deem this is so basically because there is not enough data in this kinematical region, especially for $\cos \theta_K^{c.m.} = 0.95$, and thus there is nothing that can control the behaviour of the models in this region. At central angles, the majority of fits with the hyperon resonances do not capture the data at the very threshold and some of the fits overestimate the first peak in the backward angles. Besides, in the fits where the $\lambda$ allows some of the hyperon resonance couplings to acquire non zero values, the information criteria (both the Akaike and the Bayesian one) tend to have larger values. This is a clear indication that these parameters (resonances) are not substantial for data description. Moreover, in the plot of the information criterion values against the $\lambda$ parameter, we see a significant drop once the hyperon resonances do not contribute. This is a very interesting observation and it strongly corroborates the claim of hyperon resonances not being important for this channel.

The results with additional $\Delta$ resonances are slightly better than the results with hyperon resonances. However, once we add the $\Delta$ resonances, i.e. we add more resonances to the core set of resonances, the $\chi^2$ gets worse. In other words, we add more free parameters and thus the model has more freedom to adapt to data but in a contradiction to a common sense the $\chi^2$ becomes larger not smaller. We again deem this to be a clear indication of unimportance of the $\Delta$ resonances for reliable data description in this channel.

\section{Discussion}

We concluded the fitting process with two distinct models. One of them, which we will refer to in the subsequent text by fit M, was achieved using solely the Minuit procedure for minimizing the $\chi^2$ (see Sect. II and Ref.~\cite{CLAS21}). The other one, which we will call fit L, is a result of using Minuit together with the LASSO technique as presented in Sec.~\ref{sec:adjusing}.

The fit M incorporates altogether 14 resonances: two kaon resonances, multiple nucleon resonances, one $\Delta$ resonance and no hyperon resonances. The latter feature is rather surprising given our experience with describing the $K^+\Lambda$ production channel where a plethora of hyperon resonances contribute in a significant way (see Ref.~\cite{Skoupil:2016ast}). The obtained couplings $g_1$ and $g_2$ listed in Table \ref{tab:BCSfit} are all reasonable and the same can be said about 
the hadronic form factor ranges $\Lambda_{\rm bgr}=0.87$ GeV and $\Lambda_{N}=1.45$ GeV, 
see \cite{Skoupil:2018vdh} for a description of these parameters. The fit produces results 
which are in a very good agreement with the cross section and beam asymmetry data. This analysis also showed that hyperon resonances do not play a key role in the description of the available data, and their inclusion results in negligible effects.

Besides the best Minuit fit, we revealed another noteworthy fit using only the Minuit procedure. We do not show its comparison with data as it is hardly distinguishable from the fit M. Its $\chi^2$ is 2.33, there are two $\Delta$ resonances and one of these, the D3 = $\Delta(1920)\:3/2^+$ one, has a significant effect on data description. When we omit this $\Delta$ state, the beam asymmetry falls and is in accordance with data only at very forward kaon angles. Moreover, as there is a hyperon resonance
S1 = $\Sigma(1660)\:1/2^+$ included, we could observe its effect on the beam asymmetry data description, which is negligible. This corroborates our observation with the fit M. The N7 = $N(1720)\:3/2^+$ state in this fit behaves in a similar way to the fit M, i.e. when we leave it out the beam asymmetry drops substantially, in some angular regions even to negative values of the beam asymmetry.

\begin{figure}[h]
\centering
\includegraphics[width=0.5\textwidth]{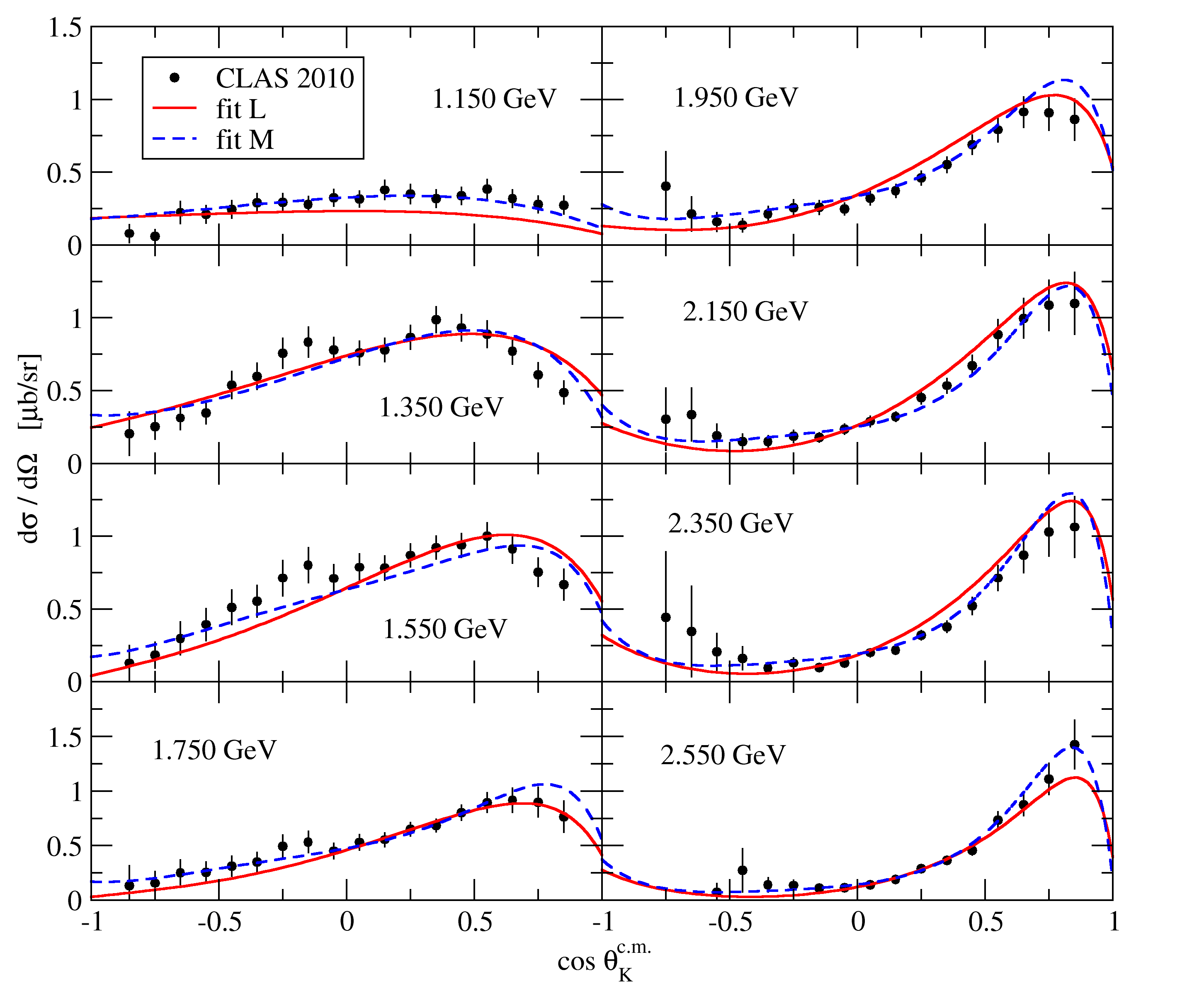}
\caption{The differential cross section as a function of the kaon center-of-mass angle $\theta_K^{c.m.}$. The points represent data from CLAS~\cite{CLAS-Pereira}. The solid red line and dashed blue lines show the models with parameters determined applying the LASSO method and Minuit alone, respectively.}
\label{fig:crs-ct0-ML}
\end{figure}

\begin{figure}[h]
    \centering
    \includegraphics[width=0.5\textwidth]{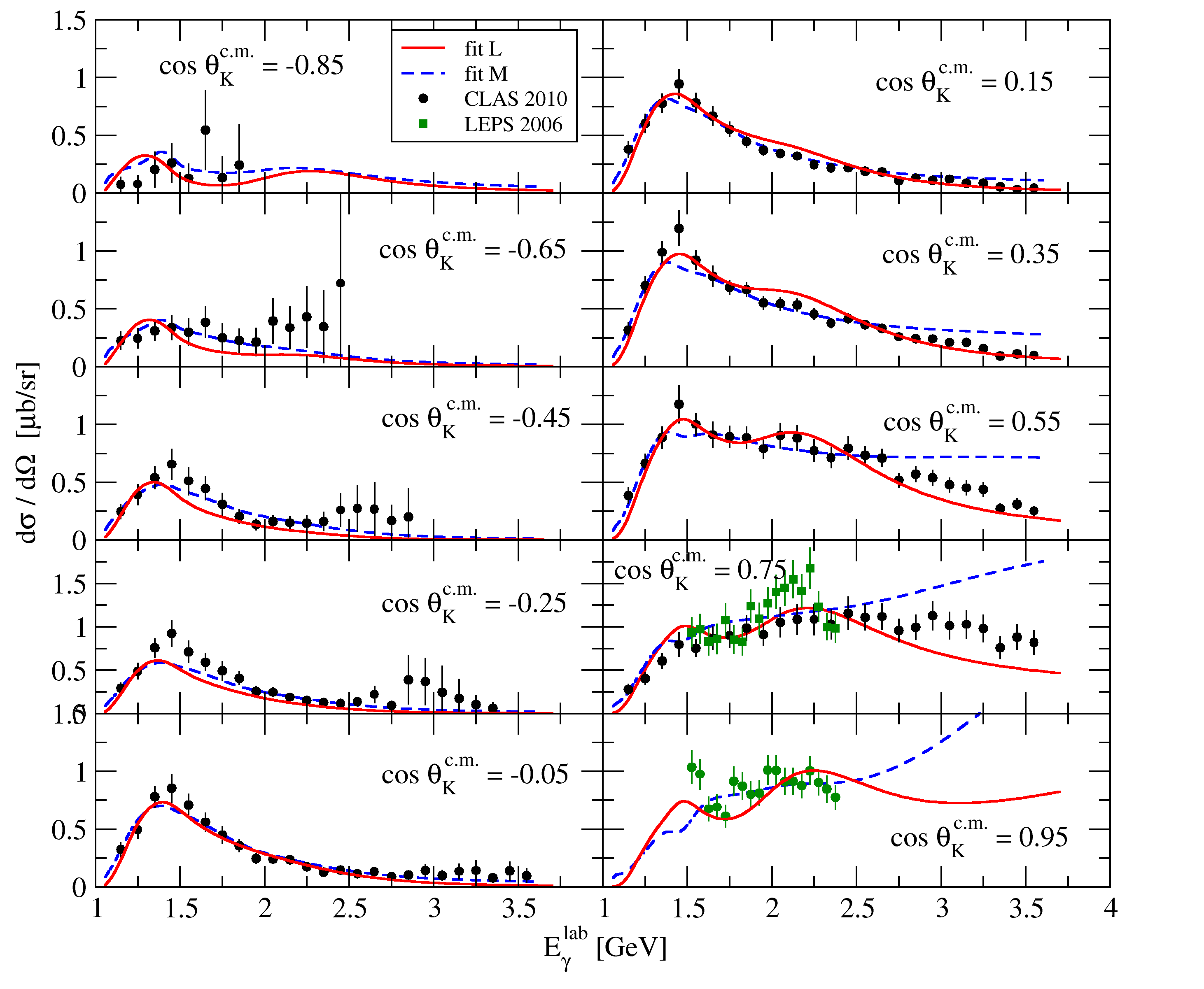}
    \caption{The differential cross section as a function of the incident photon energy $E_\gamma^{lab}$. CLAS and LEPS data are shown with the black and green points, respectively. The curves indicate the two methods used for obtained the best description of the data as described in Fig.~\ref{fig:crs-ct0-ML}.}
    \label{fig:crs-eglab-ML}
\end{figure}

\begin{figure}
    \centering
    \includegraphics[width=0.5\textwidth]{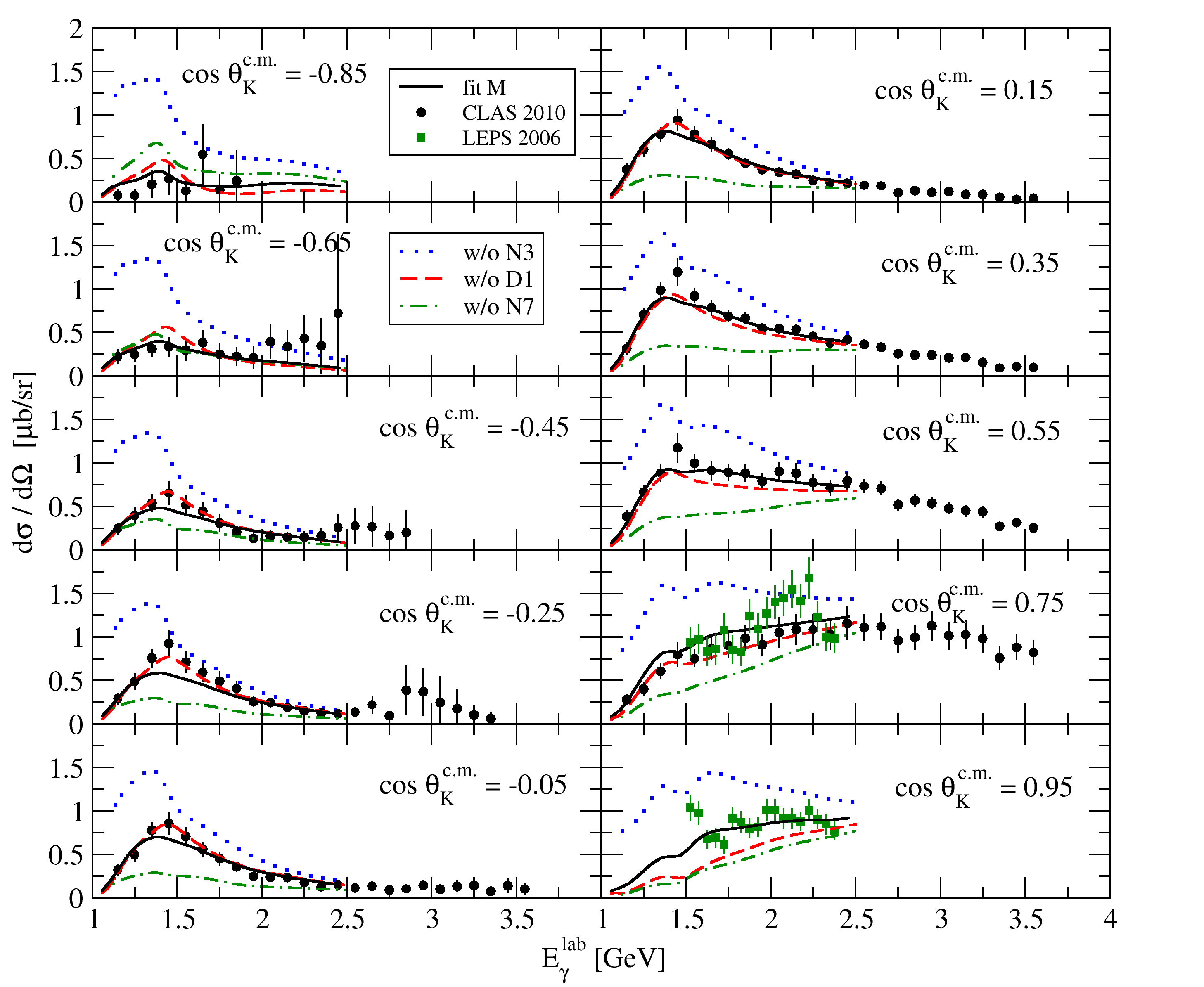}
    \caption{The differential cross section as described by the full fit M (solid line) and by the same fit with the N3 (dotted line), D1 (dashed line), and N7 (dash-dotted line) resonances omitted. The data are from CLAS~\cite{CLAS-Pereira} and LEPS~\cite{LEPS-Kohri} experiments.}
    \label{fig:crsMinuit}
\end{figure}

\begin{figure}
    \centering
    \includegraphics[width=0.5\textwidth]{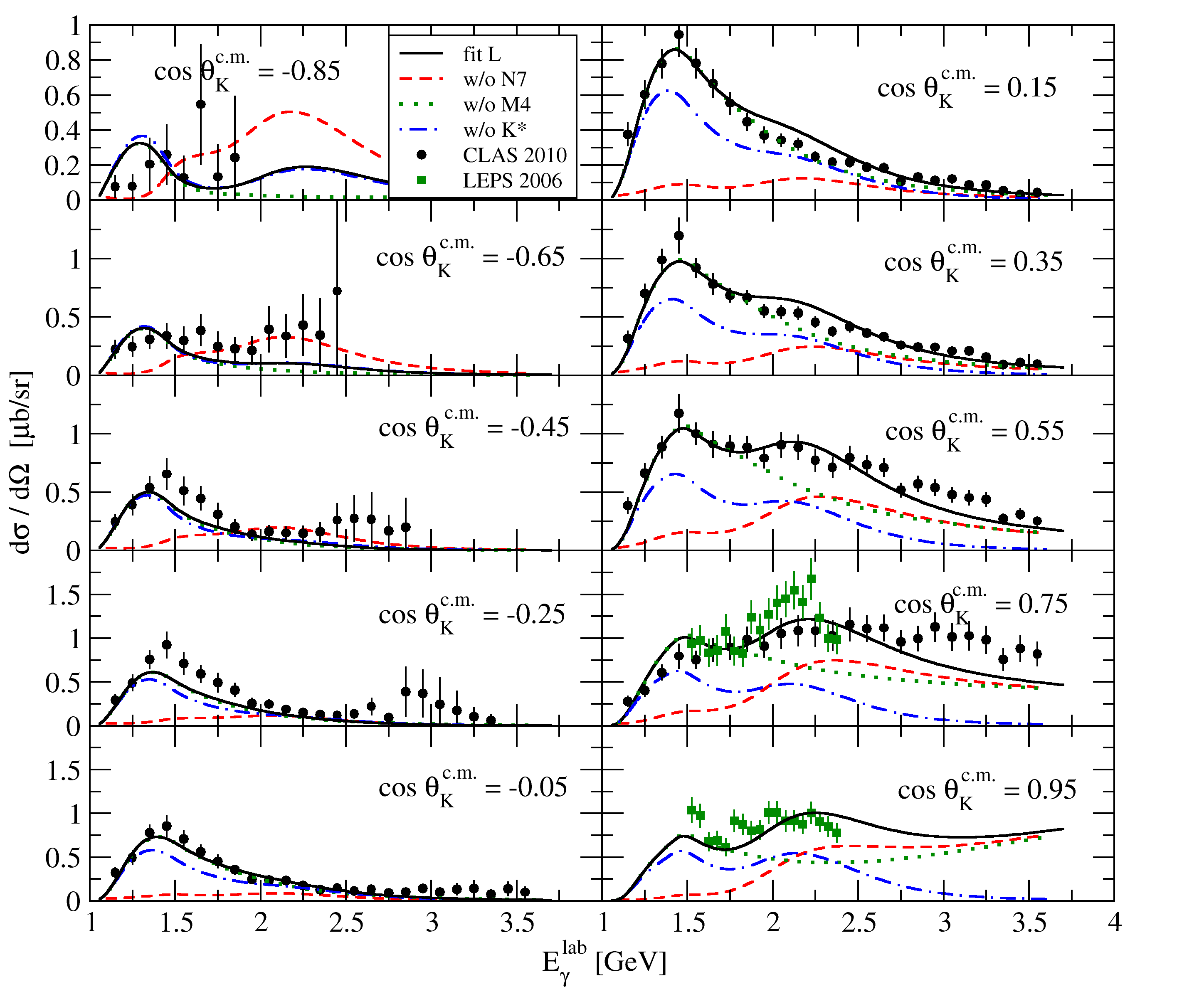}
    \caption{The differential cross section as described by the full fit L (solid line) and by the same fit with the N7 (dashed line), M4 (dotted line), and K* (dash-dotted line) resonances omitted. The data are from CLAS~\cite{CLAS-Pereira} and LEPS~\cite{LEPS-Kohri} experiments.}
    \label{fig:crsLASSO}
\end{figure}

\begin{figure}[h]
    \centering
    \includegraphics[width=0.5\textwidth]{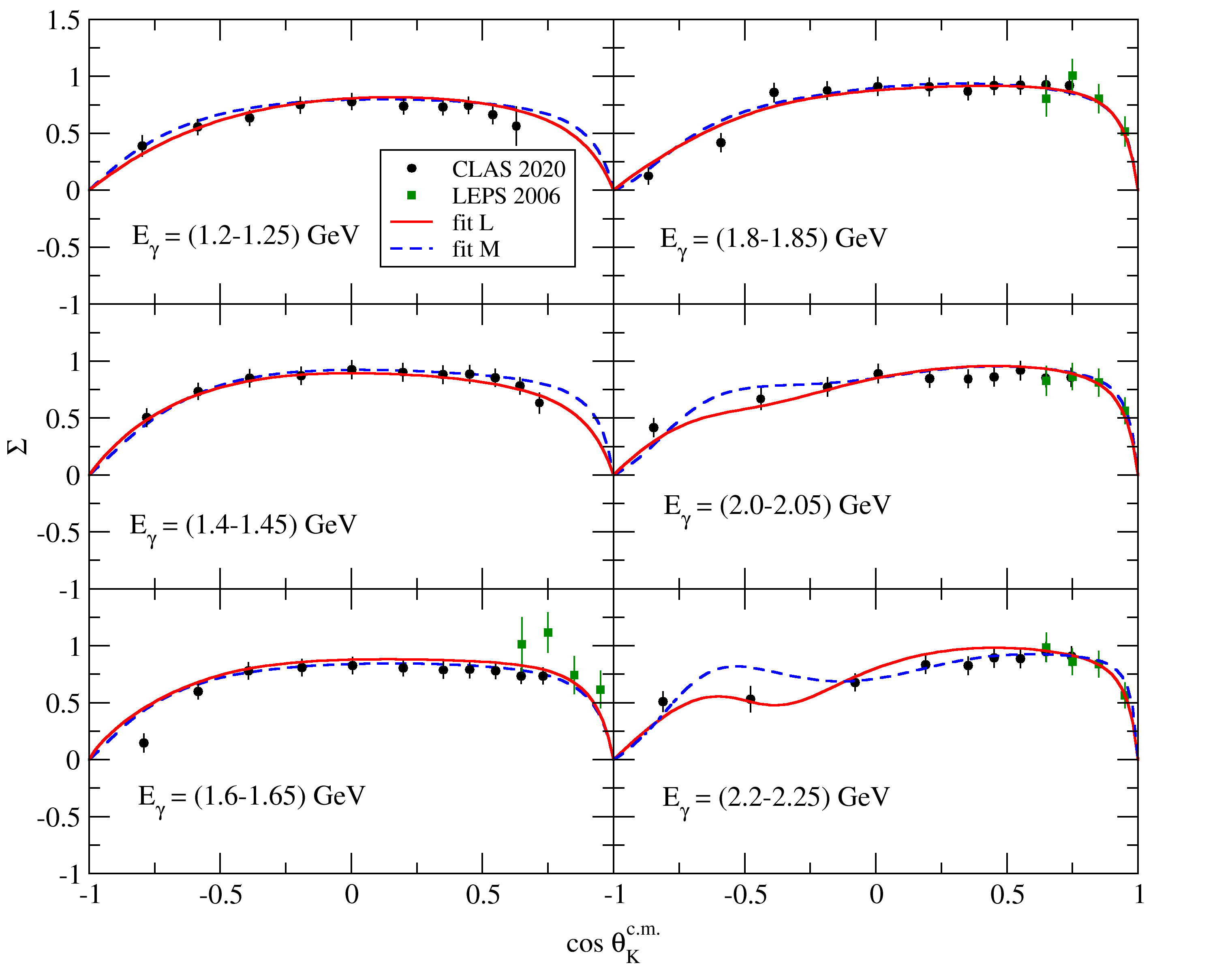}
    \caption{The photon beam asymmetry for several energetic bins in dependence on the kaon center-of-mass angle $\theta_K^{c.m.}$ as calculated by the best fit achieved by the Minuit alone and by the LASSO method and compared with the CLAS~\cite{CLAS21} data. Notation of the curves is the same as in Fig.~\ref{fig:crs-ct0-ML}.}
    \label{fig:sigma-ct0-ML}
\end{figure}

In comparison to the fit M, the $\chi^2$ of the fit L, $\chi^2 = 3.42$ is significantly larger. The large $\chi^2$ value is the price we had to pay for a smaller number of parameters. In the fit L, there are mere 17 parameters and 9 resonances while in the fit M there are 25 parameters and 14 resonances, see Table I.

In Figure~\ref{fig:crs-ct0-ML}, we compare the two best fits with one another and with the experimental data on differential cross sections from the CLAS Collaboration. The overall trend of the data is captured by both models. The fit M produces a slightly sharper peak at forward angles and for photon lab energies around 2~GeV overshoots the data, while the fit L is more moderate and tends to underestimate the data at highest photon lab energies shown. Unfortunately, neither of the models is able to capture the two-peak structure of the differential cross section above the threshold at energies 1.35 and 1.55~GeV as both models produce a smooth differential cross section at central kaon angles.

The differential cross sections in dependence on the photon lab energy $E_\gamma^{lab}$ are shown in Fig.~\ref{fig:crs-eglab-ML}. We again compare our best fits with the experimental data from the CLAS and LEPS Collaborations. In all angular bins, the models are in a satisfactory agreement with the data. The fit M tends to diverge quickly at very forward angles, while the fit L, on the other hand, underestimates the data at energies above 2.5~GeV in the $\cos \theta_K^{c.m.}=0.75$ bin. Noteworthy are also the structures that the fit L produces at forward angles. Whereas the fit M produces more or less flat cross sections, the fit L shows two broad peaks which are also supported by the data.

Among the set of included nucleon resonances in the fit M, the most noteworthy is the contribution of the $N(1720)\:3/2^{+}$ nucleon resonance whose omission leads to substantially decreased cross sections. An important effect of this resonance was also found in the $K^0\Sigma^+$ channel~\cite{Mart00}. We also note a significant contribution of the $N(1895)\:1/2^-$ state not considered in previous photoproduction analyses, with a relatively large $K\Sigma$ branching ratio. The role of the $\Delta$ resonance is in modelling the peak in the cross-section data, but it slightly modifies the beam asymmetry description as well. This can be seen once a contribution of this resonance is switched off (see Fig.~\ref{fig:crsMinuit}).

In the fit L, the most important contributions stem from the $K^*$, N7, and M4 resonances. The kaon resonance $K^*$ helps to capture the experimental data predominantly at forward angles. When we omit this resonance, the cross section falls substantially (the smaller the kaon angle, the more notable the cross-section drop becomes). The N7 nucleon resonance on its own creates the first peak and near this peak clearly dominates the model. Once we omit this resonance, the fit produces a plateau instead of a peak. On the contrary, the M4 nucleon resonance contributes in a substantial way to the description of the second peak which is more tangible at small kaon angles. Besides these three resonances, no other resonance can produce such a substantial effect in the differential cross sections (see Fig~\ref{fig:crsLASSO}). 

As the real reason why we began investigating the $K^+\Sigma^-$ channel was the new data on the photon beam asymmetries, this paper would not be complete without showing the results for this observable, which we could achieve with mere Minuit fitting procedure and also with the LASSO method. Fig.~\ref{fig:sigma-ct0-ML} shows the photon beam asymmetry in dependence on the cosine of the kaon center-of-mass angle $\theta_K^{c.m.}$. The recent CLAS data~\cite{CLAS21}  have a distinctive shape, they are large, positive and almost uniform for central kaon angles and gradually decrease at backward kaon angles. At forward kaon angles, the data are typically slightly larger than the data at backward angles (this is more notable for higher energies). The figure is complemented with the LEPS data~\cite{LEPS-Kohri} which are limited to forward going kaons and have slightly larger uncertainties than the CLAS data. Above the threshold, both fits are hardly distinguishable as both of them produce beam asymmetry which is large and positive around central kaon angles and fall off gradually at backward angles and rather abruptly at forward angles, i.e. both models can capture the experimental data well. The sole difference can be seen at energies above 2~GeV around $\cos \theta_K^{c.m.}=-0.5$ where the fit M creates a peak while the fit L shows a dip. The latter behavior is more in concert with experimental data which indicate a plateau or a dip rather than a peak in this kinematical region. 

\begin{figure}[h]
    \centering
    \includegraphics[width=0.5\textwidth]{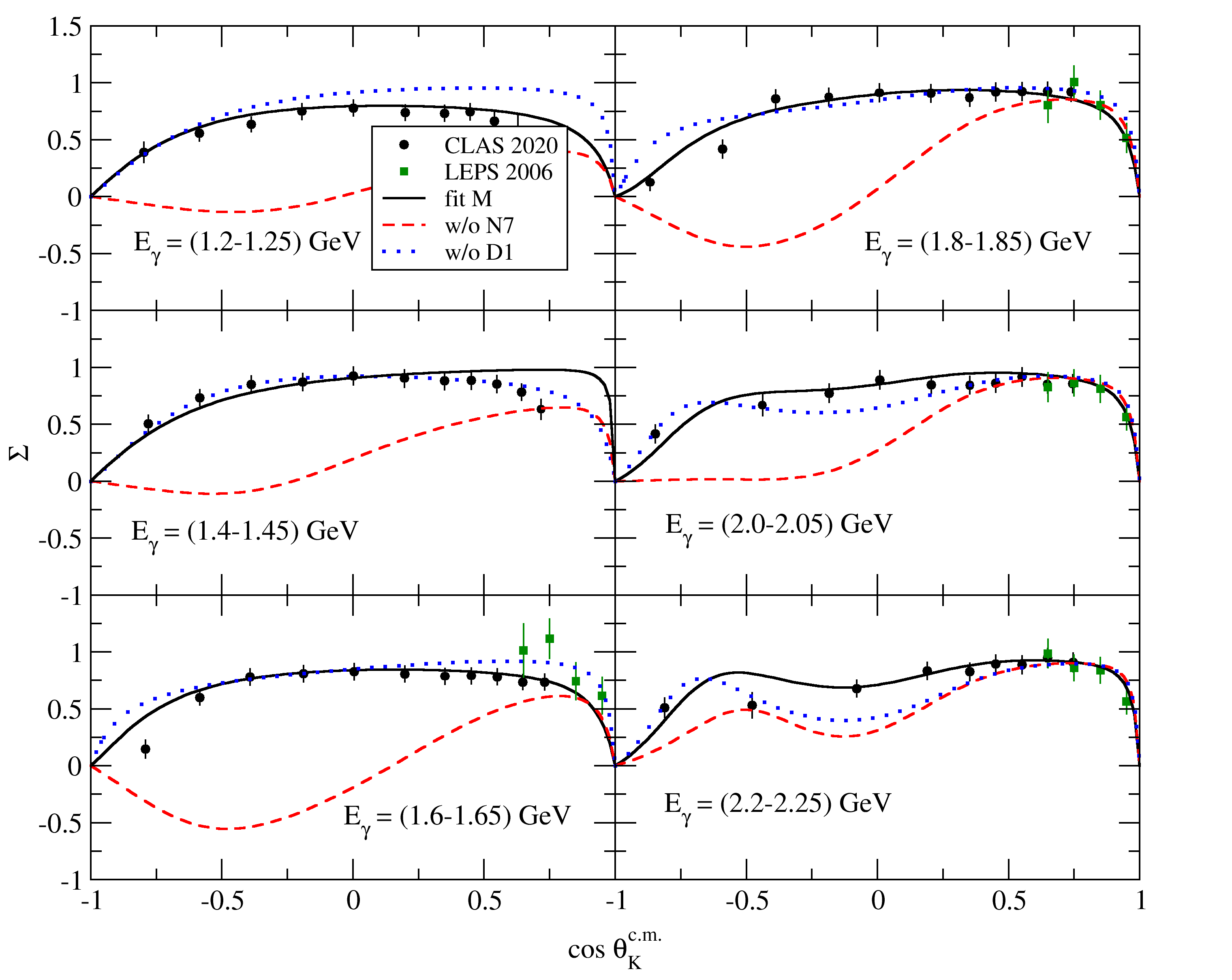}
    \caption{The photon beam asymmetry described by the full fit M (solid line) and by the same fit with the N7 (dashed line) an D1 (dotted line) resonances omitted. The data are from CLAS~\cite{CLAS21} and LEPS~\cite{LEPS-Kohri} experiments.}
    \label{fig:sigma-M}
\end{figure}

\begin{figure}[h]
    \centering
    \includegraphics[width=0.5\textwidth]{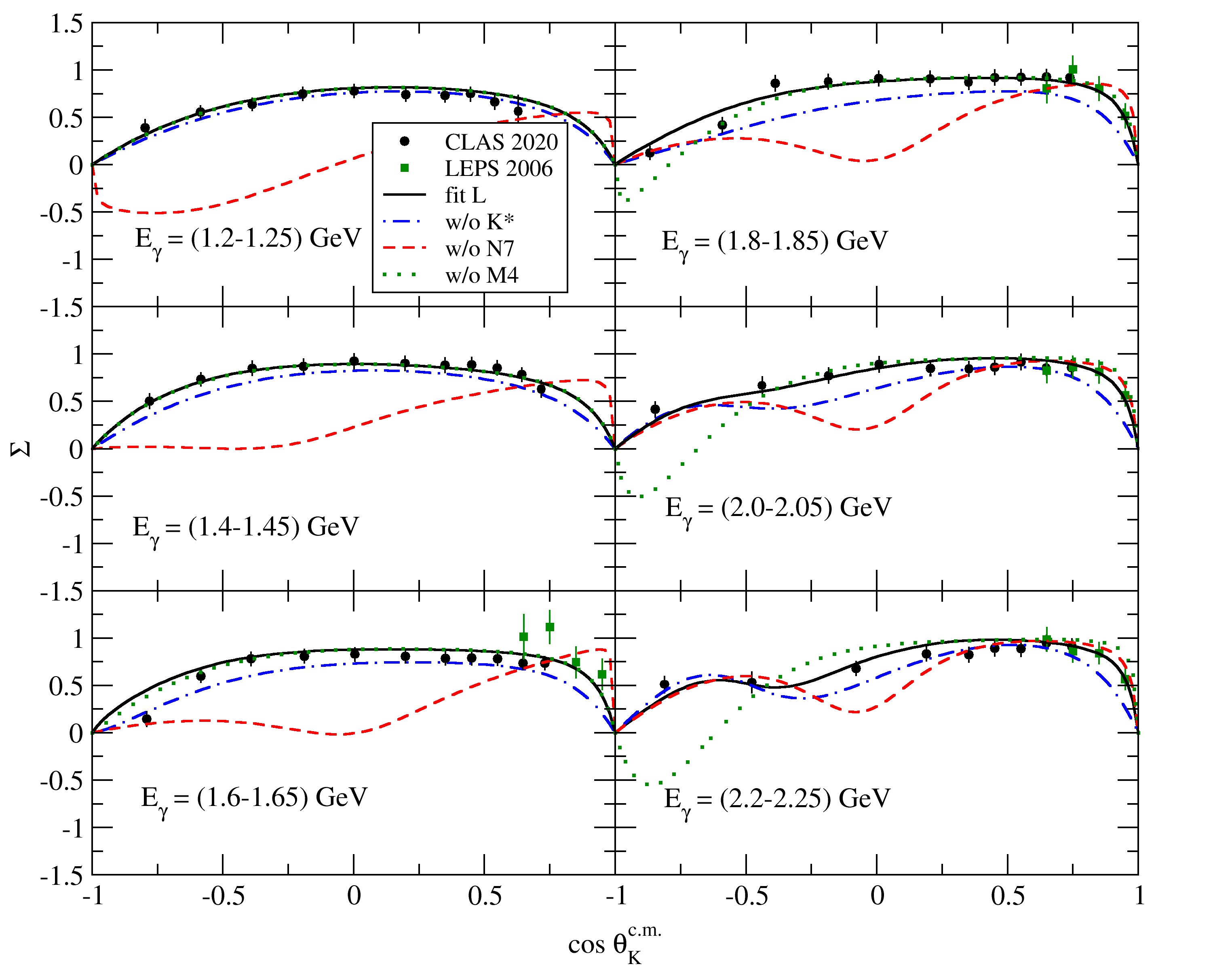}
    \caption{The photon beam asymmetry described by the full fit L (solid line) and by the same fit with the K* (dash-dotted line), N7 (dashed line), and M4 (dotted line) resonances omitted. The data are from CLAS~\cite{CLAS21} and LEPS~\cite{LEPS-Kohri} experiments.}
    \label{fig:sigma-L}
\end{figure}

In Figures~\ref{fig:sigma-M} and \ref{fig:sigma-L}, we show description of the photon beam asymmetry data by both our fits with some of the most contributing resonances omitted. The strongest contributions to the fit M, see Fig.~\ref{fig:sigma-M}, stem from N7 and D1 resonances. The omission of the N7 nucleon resonance is tangible in all energy bins as it leads to a significant drop of the photon beam asymmetry to zero and in some cases even below zero. When we leave out the D1 $\Delta$ resonance, on the other hand, we observe only minor modifications and in most energy bins we can still find agreement with data.

In the case of the fit L, we identified the $K^*$, N7, and M4 resonances to be the ones that contribute most to the photon beam asymmetry, see Fig.~\ref{fig:sigma-L}. When we omit the $K^*$ kaon resonance, we cannot capture the magnitude of the data as the model outcomes lie below the data in almost all energy bins. Similar but more pronounced outcomes are produced when we turn off the N7 nucleon resonance: Near the energy threshold, the photon beam asymmetry even changes its sign. The omission of the M4 nucleon resonance leads to a minor corrections up to $E_\gamma^{lab}=1.8\,\text{GeV}$, while beyond this energy we observe a dip at backward angles.

\begin{figure}[h]
    \centering
    \includegraphics[width=0.5\textwidth]{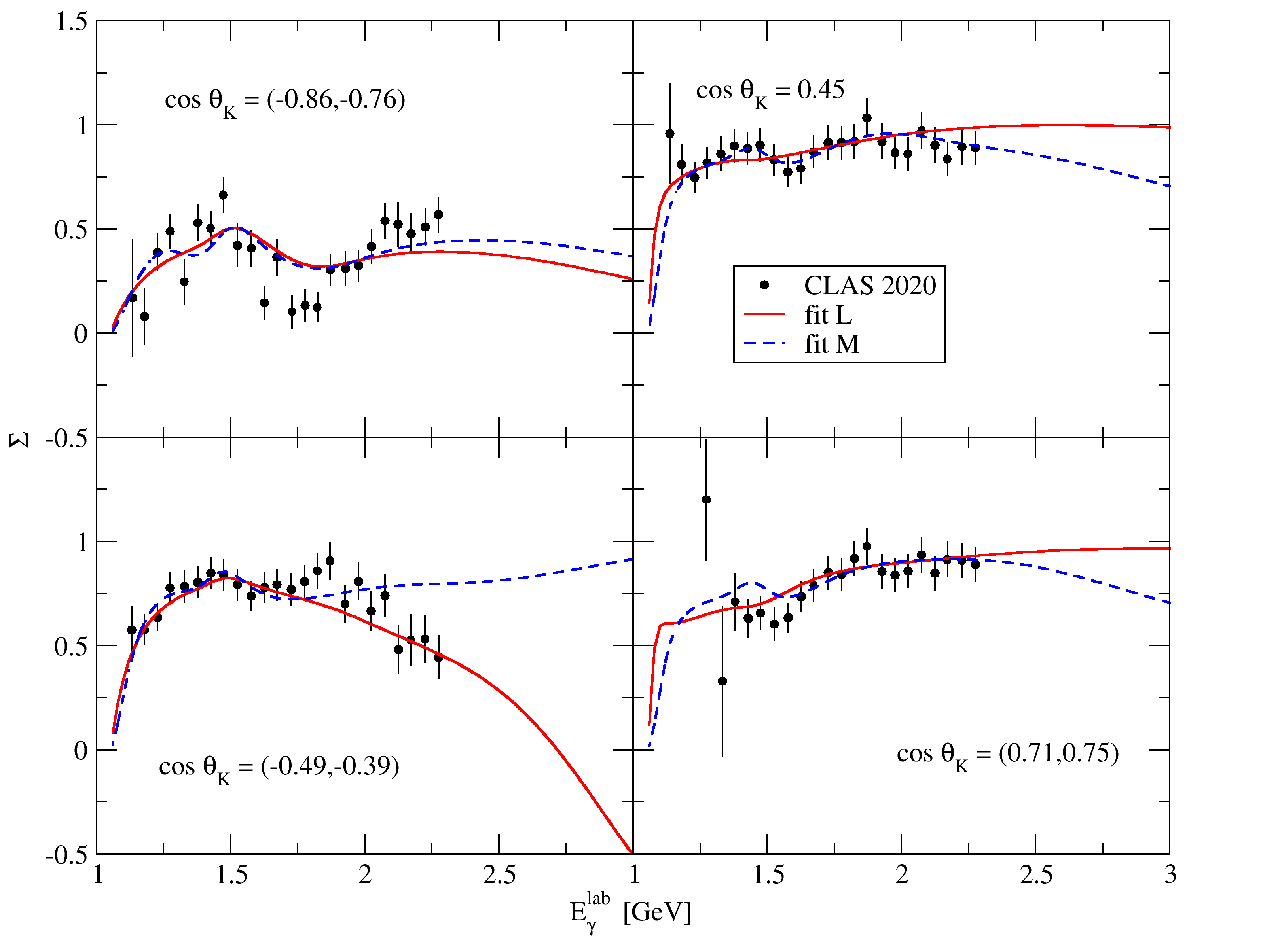}
    \caption{The photon beam asymmetry in dependence on the photon laboratory energy $E_\gamma^{lab}$ as calculated by the best fit achieved by the Minuit alone and by the LASSO method and compared with the CLAS~\cite{CLAS21} data. Notation of the curves is the same as in Fig.~\ref{fig:crs-ct0-ML}.}
    \label{fig:sigma-eglab-ML}
\end{figure}

\begin{figure}[h]
    \centering
    \includegraphics[width=0.5\textwidth]{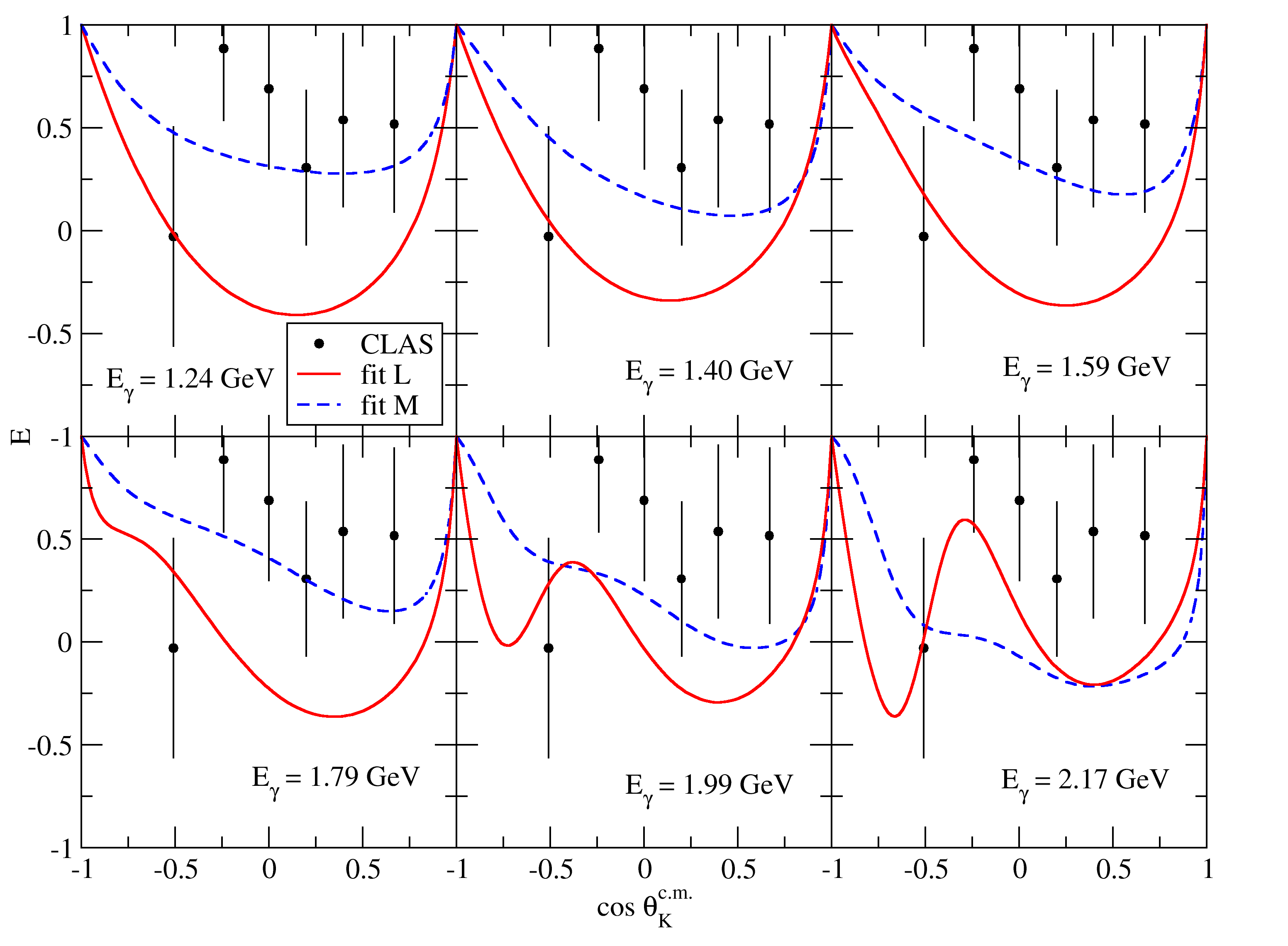}
    \caption{The CLAS beam-target asymmetry data~\cite{CLAS-E} in dependence on $\cos \theta_K^{c.m.}$ compared with predictions of the fit L (solid line) and fit M (dashed line). }
    \label{fig:e}
\end{figure}

We show also the energy dependence of the photon beam asymmetry for four angular bins, see Fig.~\ref{fig:sigma-eglab-ML}. Both fits produce again very similar results, the only difference being the behavior at energies above 2~GeV, which is most notable in the $\cos \theta_K^{c.m.} =(-0.49,-0.39)$ angular bin. Please note that the model results are calculated for the middle value in each angular bin whereas the data are scattered within the whole bin.

In Fig.~\ref{fig:e}, we show the angular dependence of the beam-target asymmetry $E$. This observable was not included among the data and thus Fig.~\ref{fig:e} illustrates merely the predictive power of our new fits. The data are positive in all energy bins shown and for energies below 1.8~GeV the fit M can capture their shape as it produces decreasing $E$. The fit L can, on the other hand, produce acceptable beam-target asymmetry only in the $E_\gamma^{lab} = 2.17~\text{GeV}$ bin; in other energy bins it underestimates all the data except for the data point at $\cos \theta_K^{c.m.}=-0.5$.

\section{Conclusion}
Exploiting the isobar model we performed an investigation of the $K^+\Sigma^-$ photoproduction off a neutron target in the resonance region. In order to reach an acceptable correspondence with experimental data, we used the tree-level Feynman diagrams with exchanges of particles in their ground as well as excited states. For high-spin nucleon states, we have used the consistent formalism where spurious lower-spin modes vanish in the amplitude. 

The cornerstone of this analysis was, however, the upgrade to the fitting method. In our previous studies, we used only the plain $\chi^2$ minimization. Such a technique, unfortunately, cannot prevent us from overfitting the data, i.e. introducing more parameters (and thus resonances) than are needed for the data description. In the third nucleon-resonance region, where the process under study in this paper occurs, there are a plenty of resonant states which overlap each other. It is therefore of crucial importance to limit the number of states which we consider in our analysis. A solution to this issue seems to be a method called regularization, where we introduce a penalty term which in effect restricts the number of nonzero parameters and in this way hinders us from overfitting the data.

With help of both plain $\chi^2$ minimization and the more advanced LASSO method, we could arrive at two models which both give us satisfactory agreement with data. We discussed the course of the fitting process, commented on the outcomes, and identified the resonant states which contribute most to the differential cross sections and photon beam asymmetries in various kinematic regions. We observe only slight differences in data description by our models, the most notable ones being the description of differential cross sections at very forward angles where the fit L produces two broad peaks whilst the fit M is flat, and the photon beam asymmetries beyond 2~GeV at backward angles where the fit M produces a bump which is not supported by the data. In both fits, the $N(1720)3/2^+$ resonance was found to be important for correct description of data.

A natural step forward for us will be using the LASSO method to perform the multi-channel analysis of the photoproduction as we deem this method to be of immense importance for selecting the optimal resonances contributing to the process.

\section{Acknowledgement}
The authors would like to thank H.~Haberzettl for drawing their attention to the LASSO method. This work was supported by the Grant Agency of the Czech Republic under Grant
No. 19-19640S (P.B, A.C. and D.S.) and 19-14048S (D.P.).
\appendix

\section{Contributions to the Invariant Amplitude}

We consider the process 
\begin{equation}
\gamma_V(k) + n (p) \rightarrow K^+(p_K) + \Sigma^-(p_\Sigma)
\end{equation}
with the corresponding four-momenta given in the parentheses; the four-momentum of the intermediate particle is denoted by $q$. In the following sections, we summarize the invariant amplitudes and include also the hadron form factors, $f_x(x), x=s,t,u$, which turn out to be essential for the gauge invariance of the full amplitude.

\subsection{Born $s$ channel}
The invariant amplitude of the proton exchange reads
\begin{equation}
\mathbb{M}_{Bs} = \bar{u}(p_\Sigma)V_S f_s(s) \frac{\not\!p + \not\! k + m_p}{s-m_p^2} V_\mu^{EM} \varepsilon^{\mu}(k)u(p),
\end{equation}
where
\begin{equation}
V_S = ig_{K\Sigma n}\gamma_5
\label{eq:strongV}
\end{equation}
and
\begin{equation}
V_{\mu}^{EM} = i\frac{\kappa_n}{2m_n}\sigma_{\mu\nu}k^\nu
\label{eq:elmagV-s}
\end{equation}
are strong and electromagnetic vertex functions, respectively, and $\sigma_{\mu\nu} = \tfrac{i}{2}[\gamma_\mu,\gamma_\nu]$.

When we recast the amplitude into the compact form,
\begin{equation}
\mathbb{M} = \bar{u}(p_\Lambda) \sum_{j=1}^{6}\mathcal{A}_j(k^2,s,t,u) \mathcal{M}_j u(p), 
\label{eq:compact}
\end{equation}
we can extract the scalar amplitudes
\begin{subequations}
\begin{align}
\mathcal{A}_1 &= \frac{g_{K\Sigma n}}{s-m_n^2}\kappa_n f_s(s), \\
\mathcal{A}_4 &= \frac{g_{K\Sigma n}}{s-m_n^2}\frac{\kappa_n}{m_n} f_s(s) = -2\mathcal{A}_6.
\end{align}
\end{subequations}
The operators $\mathcal{M}_j$ appearing in Eq.~(\ref{eq:compact}) are defined in Eqs.~(17) in Ref.~\cite{Skoupil:2016ast}.

\subsection{Born $t$ channel}
The invariant amplitude of the kaon exchange in the $t$ channel reads
\begin{equation}
\mathbb{M}_{Bt} = \bar{u}(p_\Sigma) V_S f_t(t) \frac{1}{t-m_K^2} V_{\mu}^{EM}\varepsilon^\mu(k)u(p)
\end{equation}
which can be cast to the compact form
\begin{equation}
\mathbb{M}_{Bt} = \bar{u}(p_\Sigma)\gamma_5\left( \mathcal{A}_2\mathcal{M}_2 + \mathcal{A}_3\mathcal{M}_3 - g_{K\Sigma n} f_t(t) \frac{k\cdot \varepsilon}{k^2}\right)u(p).
\end{equation}
The last term in the brackets is a gauge-invariance breaking term. There are only two nonzero scalar amplitudes
\begin{equation}
\mathcal{A}_2 = -\mathcal{A}_3 = 2\frac{g_{K\Sigma n}}{t-m_K^2} f_t(t).
\end{equation}

\subsection{Born $u$ channel}
The electromagnetic $\gamma \Sigma \Sigma$ vertex factor has the form
\begin{equation}
V_{\mu}^{EM} = \gamma_\mu + i\frac{\kappa_\Sigma}{2m_\Sigma}\sigma_{\mu\nu} k^\nu
\end{equation}
and the strong vertex function is the same as in Eq.~(\ref{eq:strongV}). The hyperon exchange in the $u$ channel reads
\begin{equation}
\mathbb{M}_{Bu} = \bar{u}(p_\Sigma)V_{\mu}^{EM}\frac{\not\!p_\Sigma - \not\! k + m_\Sigma}{u-m_\Sigma^2}V_S f_u(u) \varepsilon^\mu(k) u(p)
\end{equation}
and can be again recast into the compact form 
\begin{equation}
\begin{aligned}
\mathbb{M}_{Bu} &= \bar{u}(p_\Sigma)\Big( \mathcal{A}_1\mathcal{M}_1 + \mathcal{A}_5\mathcal{M}_5 + \mathcal{A}_6\mathcal{M}_6\\
& \,\,\,\, + g_{K\Sigma n} f_u(u) \frac{k\cdot \varepsilon}{k^2} \Big) u(p).
\end{aligned}
\end{equation}
Similarly to the $t$-channel exchange, the last term in the brackets is the gauge-invariance breaking term. In case we do not assume hadron form factors, i.e. $f_t = f_u = 1$, the gauge-invariance terms in $t$ and $u$ channels cancel each other and the resulting amplitude is gauge invariant. 

The scalar amplitudes are 
\begin{subequations}
\begin{align}
\mathcal{A}_1 &= \frac{g_{K\Sigma n}}{u-m_\Sigma^2} f_u(u),\\
\mathcal{A}_5 &= \frac{g_{K\Sigma n}}{u-m_\Sigma^2}\frac{\kappa_\Sigma}{m_\Lambda} f_u(u) = 2\mathcal{A}_6.
\end{align}
\end{subequations}

\subsection{Contact current}
In case we introduce the hadron form factors, the full amplitude contains gauge noninvariant terms. In order to get rid of them, we consider a contact term which acquires the form
\begin{equation}
\begin{aligned}
\mathbb{M}_{cc} &= \bar{u}(p_\Sigma) V_S e \bigg[ -(2p_K-k)^\mu \frac{f_t-1}{t-m_K^2} f_u \\
& + (2p_\Sigma -k)^\mu \frac{f_u-1}{u-m_\Sigma^2}f_t\bigg] \varepsilon_\mu (k) u(p)
\end{aligned}
\end{equation}
and can be recast into the compact form
\begin{equation}
\begin{aligned}
\mathbb{M}_{cc} &= ieg_{K\Sigma n} u(p)\gamma_5 \bigg\{ \bigg[ -2\mathcal{M}_2+2\mathcal{M}_3 \\
& + (t-m_K^2)\frac{k\cdot \varepsilon}{k^2} \bigg] \frac{f_t -1 }{t-m_K^2}f_u \\
& + \bigg[2\mathcal{M}_3 - (u-m_\Sigma^2) \frac{k\cdot \varepsilon}{k^2} \bigg] \frac{f_u-1}{u-m_\Sigma^2} f_t \bigg\} u(p).
\end{aligned}
\end{equation}

Its contributions to the scalar amplitudes read
\begin{subequations}
\begin{align}
\mathcal{A}_2 &= - 2ieg_{K\Sigma n}  \frac{f_t -1 }{t-m_K^2}f_u,\\
\mathcal{A}_3 &= 2ieg_{K\Sigma n} \bigg[ \frac{f_t -1 }{t-m_K^2}f_u + \frac{f_u-1}{u-m_\Sigma^2} f_t \bigg].
\end{align}
\end{subequations}
and the gauge-invariance breaking terms abolish those terms in the $t$ (A7) and $u$ (A11) channel exchanges.

\section{Aspects of the fitting procedure}
\label{sec:app-reg}

\subsection{Ridge and LASSO regularization}

Typically, the regularization term is a norm of the parameter vector $\bm{\theta} = (\theta_1,\dots,\theta_n)$ multiplied by a regularization parameter $\lambda$ that determines the amount of the penalty on the parameter values
\begin{equation}
P(\lambda) = \lambda \sum_{j=1}^n |\theta_j|^q
\end{equation}

The two most common types of regularization correspond to $q=1$ and $q=2$. The $q=1$ case, or L$_1$-norm regularization, is also known as LASSO \cite{Lasso}, while the $q=2$ case, or L$_2$-norm regularization, is also known as Ridge \cite{Ridge}. Minimizing the new function $\chi_T^2 = \chi^2 + P(\lambda)$ is equivalent to  minimizing $\chi^2$, subject to the constraint $\sum_{j=1}^n |\theta_j|^q \leq c$ (where $c > 0$). Due to the geometry of the constraint in parameter space (hyperoctahedron for LASSO, as opposed to hypersphere for Ridge) LASSO forces some of the parameters to zero, thus favoring a sparser model. This characteristic of LASSO makes it more suitable for feature selection and this is why we use it in our approach.


\subsection{Parameter selection procedure}

In machine learning, a model's accuracy is not assessed by the error calculated on the \textit{training set}, which is used to fit its parameters, but on an independent \textit{test set} that is intentionally held out of the original dataset. In the process called $k$-fold cross-validation, the original dataset is partitioned in $k$ samples of equal size, each of which serves as the test set against which the model trained on the rest of the data (the remaining $k-1$ samples) will be evaluated and the results of these runs are then averaged \cite{Bishop}.   
 
When more parameters are added to a model, the training set error will invariably decrease, but if the test set error increases it is a sign that the model is overfitted to the training set and fails to generalize to new data. So, using only a subset of a model's parameters may improve its prediction accuracy, as well as its interpretability.

Choosing the best subset can be done through either forward or backward stepwise selection \cite{ESL}. Forward selection consists in sequentially adding the parameter that most improves the fit, while backward selection consists in sequentially removing the parameter that, when removed, least worsens the fit. In both cases, it is the test set (prediction) error that decides which is the best parameter subset. Forward selection has a wider applicability than backward selection, as it can be used also in cases where the number of parameters exceeds the number of data points, while backward selection cannot be used in such cases. Naturally, these selection procedures become highly impractical when a large number of parameters is involved, resulting in extremely large number of subsets to be evaluated.
 
In the present work the parameter subset selection is done automatically by changing the regularization parameter $\lambda$ accordingly; e.g. to implement forward selection we start from a large value of $\lambda$ that leads to a very sparse model and gradually decrease it to zero, leading to the full unregularized model. The optimal value of $\lambda$ and hence the best parameter subset, is chosen based on information criteria which are commonly used in model selection. Furthermore, it has been shown that model selection by cross-validation is asymptotically equivalent to Akaike's information criterion \cite{Stone}.


\subsection{Maximum likelihood estimation}

In a measurement process, an observation $d_i$  can be represented by a random variable $D_i$ characterized by a probability density function $f(D_i|\theta)$, which depends on a set of parameters $\bm{\theta}=(\theta_1,\dots,\theta_n)$. The joint probability of N independent and identically distributed measurements is given by the product of the individual densities
\begin{equation}
\label{l1}
P(\bm{D}|\bm{\theta}) = P(D_1,\dots,D_N|\bm{\theta}) = \prod_{i=1}^{N} f(D_i|\bm{\theta}).
\end{equation}
For a sample of data $\bm{d} = (d_1,\dots,d_N)$, this joint probability distribution (now only a function of $\bm{\theta}$) defines the likelihood function
\begin{equation}
\label{l2}
L_{\bm{d}}(\bm{\theta}) = P(D=\bm{d}|\bm{\theta})
\end{equation}
for that particular sample and expresses how probable the observed data $\bm{d}$ is, for given values of the parameters $\bm{\theta}$. In maximum likelihood estimation (MLE), we seek the parameter values $\bm{\hat{\theta}}$ that maximize the likelihood, $(L_{\bm{d}})_{\mathrm{max}} = L_{\bm{d}}(\bm{\hat{\theta}})$ and hence the probability of observing the specific sample of data. For computational reasons, it is the natural logarithm of the likelihood, $\ln L_{\bm{d}}(\bm{\hat{\theta}})$, that is commonly maximized.
Furthermore, the fundamental assumption is made that the outcome $d_i$ of each  measurement follows a Gaussian distribution characterized by a given variance $\sigma_i^2$ (given by experiment) and a mean value $\mu_i$. 
In fitting a theoretical model that depends on a set of parameters $\bm{c}=(c_1,\dots,c_n)$ to the data, we adjust the parameters of the model so that its predictions $p_i(\bm{c})$ provide the mean values that maximize the likelihood of the particular sample. The likelihood function to be maximized thus becomes
\begin{equation}
\label{l3}
L_{\bm{d}}(\bm{c}) = \prod_{i=1}^{N}(2\pi\sigma_i^2)^{-1/2}\exp \left( -\frac{(d_i-p_i(c))^2}{2\sigma_i^2}\right)
\end{equation}
and the log-likelihood
\begin{equation}
\label{l4}
\ln L_{\bm{d}}(\bm{c}) = \mathrm{const} - \sum_{i=1}^{N}\frac{(d_i-p_i(c))^2}{2\sigma_i^2} = \mathrm{const} - \chi^2
\end{equation}
Thus, under the above assumptions, maximizing the log-likelihood is equivalent to minimizing $\chi^2$.
\subsection{Derivations of information criteria}
The following two sections present the basic steps in deriving the Akaike (AIC) and the Bayesian (BIC) information criteria and are based on \cite{BA} and \cite{ICSM, Bishop} respectively.
\subsubsection{Akaike Information Criterion}
The Akaike Information Criterion \cite{Akaike} is an extension of the maximum likelihood principle based on the notion of relative entropy, or Kullback-Leibler (K-L) divergence, from information theory \cite{Cover}.

The K-L divergence is defined as
\begin{equation}
\label{KLdef}
I(f,g) = \int f(x) \ln\left(\frac{f(x)}{g(x|\bm{\theta})}\right)dx \equiv E_x\left[\ln\left(\frac{f(x)}{g(x|\bm{\theta})}\right)\right]
\end{equation}
and expresses the degree of dissimilarity between the true (but unknown) probability distribution $f(x)$ that generates the data and an approximating distribution $g(x|\bm{\theta})$ that is specified by a set of parameters $\bm{\theta} = (\theta_1, \theta_2,...,\theta_n)$. It can also be regarded as the expected value of $\ln\left(f(x) / g(x|\bm{\theta})\right)$ with respect to the true distribution $f(x)$, for the whole population $x$.
Using the likelihood function in place of $g(x|\bm{\theta})$, one observes that the value $\bm{\theta_0}$ that maximizes likelihood also minimizes the K-L divergence, meaning that $g(x|\bm{\theta_0})$ is as close as possible to the true $f(x)$. 

However, since our data $\bm{d}$ is a sample taken from the statistical population $x$, we can only make an estimate $\bm{\hat{\theta}}(\bm{d})$ of the true $\bm{\theta_0}$, based on this sample. Hence, the aim of seeking the minimum $I(f,g(\cdot|\bm{\theta_0}))$ is replaced by finding the (larger) minimum of the average $E_{\bm{d}}[I(f,g(\cdot|\bm{\hat{\theta}(d})))]$ over repeated samples $\bm{d}$. It should be noted that with increasing sample size the likelihood maximizing $\bm{\theta}_{\mathrm{MLE}} = \bm{\hat{\theta}}$ approaches the true $\bm{\theta_0}$.

For $\bm{\theta} = \bm{\hat{\theta}}(\bm{d})$, Eq.~(\ref{KLdef}) can be written as
\begin{equation}
\label{I2}
\begin{split}
I(f,g(\cdot|\bm{\hat{\theta}}(\bm{d}))) &= \int f(x) \ln f(x) dx - \int f(x) \ln g(x|\bm{\hat{\theta}}(\bm{d}))dx\\
&= \text{constant} - E_x[\ln g(x|\bm{\hat{\theta}}(\bm{d}))]
\end{split}
\end{equation}
where the first term does not depend on $\bm{\theta}$, so the focus will be on $E_x[\ln g(x|\bm{\hat{\theta}}(\bm{d}))]$.

If we Taylor-expand $\ln g(x|\bm{\hat{\theta}})$ around $\bm{\theta_0}$ up to second order we get
\begin{equation}
\label{log1}
\begin{split}
\ln g(x|\bm{\hat{\theta}}) \approx & \ln g(x|\bm{\theta_0}) + (\bm{\hat{\theta}} - \bm{\theta_0})^\intercal \nabla \ln g(x|\bm{\theta_0})\\ 
&+ \frac{1}{2} (\bm{\hat{\theta}} - \bm{\theta_0})^\intercal \nabla^2 \ln g(x|\bm{\theta_0})(\bm{\hat{\theta}} - \bm{\theta_0})
\end{split}  
\end{equation}
with the 2nd term containing the gradient vector and the 3rd term the Hessian matrix of second derivatives, both evaluated at $\bm{\theta_0}$. Taking the expected value $E_x[\dots]$ of both sides of Eq.~(\ref{log1}), as defined in Eq.~(\ref{KLdef}), it is easily shown that the gradient term vanishes since $\bm{\theta_0}$ is the true minimum of $I(f,g(\cdot|\bm{\theta}))$. So, from Eq.~(\ref{log1}) we have
\begin{equation}
\label{log2}
E_x[\ln g(x|\bm{\hat{\theta}})] \approx E_x[\ln g(x|\bm{\theta_0})] - \frac{1}{2} (\bm{\hat{\theta}} - \bm{\theta_0})^\intercal \mathrm{I}(\bm{\theta_0})(\bm{\hat{\theta}} - \bm{\theta_0})
\end{equation} 
where $\mathrm{I}(\bm{\theta_0})$ is the $n \times n$ matrix with matrix elements
\begin{equation}
\mathrm{I}(\bm{\theta_0})_{ij} = E_x\left[-\left.\frac{\partial^2 \ln g(x|\bm{\theta})}{\partial \theta_i \partial \theta_j}\right\rvert_{\bm{\theta} = \bm{\theta_0}}\right]
\end{equation}
known as Fisher information matrix.
As already mentioned, the quantity we seek to minimize is the average $E_{\bm{d}}[I(f,g(\cdot|\bm{\hat{\theta}(d})))]$ over different samples $\bm{d}$, which can be written as $E_{\bm{\hat{\theta}}}[I(f,g(\cdot|\bm{\hat{\theta}}))]$ i.e. as the average over the different $\bm{\hat{\theta}}$'s of each sample.

Therefore, the relevant quantity to maximize (because of the minus sign in Eq.~(\ref{I2})) is 
\begin{equation}
\label{T0}
T \equiv E_{\bm{\hat{\theta}}}E_x[\ln g(x|\bm{\hat{\theta}})] 
\end{equation}
Using Eq.~(\ref{log2}), $T$ can be written
\begin{equation}
\label{T1}
T \approx E_x[\ln g(x|\bm{\theta_0})] - \frac{1}{2} E_{\bm{\hat{\theta}}}[(\bm{\hat{\theta}} - \bm{\theta_0})^\intercal \mathrm{I}(\bm{\theta_0})(\bm{\hat{\theta}} - \bm{\theta_0})]
\end{equation}   
since the first term does not depend on $\bm{\hat{\theta}}$. Taking into account the matrix identity involving the trace $z^\intercal A z = \mathrm{tr} \left( A z z^\intercal \right)$ and the fact that $I(\bm{\theta_0})$ is independent of $\bm{\hat{\theta}}$, Eq.~(\ref{T1}) becomes
\begin{equation}
\label{T2}
T \approx E_x[\ln g(x|\bm{\theta_0})] - \frac{1}{2} \mathrm{tr}\left( \mathrm{I}(\bm{\theta_0})E_{\bm{\hat{\theta}}}[(\bm{\hat{\theta}} - \bm{\theta_0})(\bm{\hat{\theta}} - \bm{\theta_0})^\intercal]\right).
\end{equation}  
In the large sample limit, $E_{\bm{\hat{\theta}}}[(\bm{\hat{\theta}} - \bm{\theta_0})(\bm{\hat{\theta}} - \bm{\theta_0})^\intercal]$ is the covariance matrix of the maximum likelihood estimate, denoted by $\Sigma$, so $T$ is written
\begin{equation}
\label{T}
T \approx E_x[\ln g(x|\bm{\theta_0})] - \frac{1}{2} \mathrm{tr}\left( \mathrm{I}(\bm{\theta_0})\Sigma\right)
\end{equation}

In the next steps, the first term, $E_x[\ln g(x|\bm{\theta_0})]$, is approximated in a similar way as $E_x[\ln g(x|\bm{\hat{\theta}})]$ in Eqs.~(\ref{log1},\ref{log2}), but with 
$\bm{\hat{\theta}}$ and $\bm{\theta_0}$ switched. So, $\ln g(x|\bm{\theta_0})$ is expanded around $\bm{\hat{\theta}}$
\begin{equation}
\label{expan}
\begin{split}
\ln g(x|\bm{\theta_0}) \approx & \ln g(x|\bm{\hat{\theta}}) + (\bm{\theta_0} - \bm{\hat{\theta}})^\intercal \nabla \ln g(x|\bm{\hat{\theta}})\\ 
&+ \frac{1}{2} (\bm{\theta_0} - \bm{\hat{\theta}})^\intercal \nabla^2 \ln g(x|\bm{\hat{\theta}})(\bm{\theta_0} - \bm{\hat{\theta}})
\end{split}  
\end{equation}
only, this time, the gradient vanishes before taking the expected value, as it is evaluated at $\bm{\hat{\theta}}$, which is the value that maximizes the log-likelihood $\ln g(x|\bm{\theta})$, leading to
\begin{equation}
\label{expan1}
\begin{split}
\ln g(x|\bm{\theta_0}) \approx & \ln g(x|\bm{\hat{\theta}}) + \frac{1}{2} (\bm{\theta_0} - \bm{\hat{\theta}})^\intercal \nabla^2 \ln g(x|\bm{\hat{\theta}})(\bm{\theta_0} - \bm{\hat{\theta}})
\end{split}  
\end{equation}
Taking the expected value with respect to $x$, Eq.~(\ref{expan1}) becomes
\begin{equation}
E_x[\ln g(x|\bm{\theta_0)}] \approx E_x[\ln g(x|\bm{\hat{\theta}})] - \frac{1}{2} E_x[(\bm{\theta_0} - \bm{\hat{\theta}})^\intercal\hat{\mathrm{I}}(\bm{\hat{\theta}})(\bm{\theta_0} - \bm{\hat{\theta}}) ]
\end{equation}
with $\hat{\mathrm{I}}(\bm{\hat{\theta}})$  the $n \times n$ matrix with matrix elements
\begin{equation}
\hat{\mathrm{I}}(\bm{\hat{\theta}})_{kl} = -\left.\frac{\partial^2 \ln g(x|\bm{\theta})}{\partial \theta_k \partial \theta_l}\right\rvert_{\bm{\theta} = \bm{\hat{\theta}}}
\end{equation}
and employing again the matrix property $z^\intercal A z = \mathrm{tr} \left( A z z^\intercal \right)$
\begin{equation}
\begin{split}
&E_x[\ln g(x|\bm{\theta_0)}]\\
&\approx E_x[\ln g(x|\bm{\hat{\theta}})] - \frac{1}{2} \mathrm{tr}\left( E_x[\hat{\mathrm{I}}(\bm{\hat{\theta}})(\bm{\theta_0} - \bm{\hat{\theta}}) (\bm{\theta_0} - \bm{\hat{\theta}})^\intercal]\right)
\end{split}
\end{equation}
For large samples, the approximation $\mathrm{I}(\bm{\theta_0})  \approx  \hat{\mathrm{I}}(\bm{\hat{\theta}})$ is valid and the expected value inside the trace becomes
\begin{equation}
\begin{split}
E_x&[\hat{\mathrm{I}}(\bm{\hat{\theta}})(\bm{\theta_0} - \bm{\hat{\theta}}) (\bm{\theta_0} - \bm{\hat{\theta}})^\intercal] \approx \mathrm{I}(\bm{\theta_0})E_x[(\bm{\theta_0} - \bm{\hat{\theta}})(\bm{\theta_0} - \bm{\hat{\theta}})^\intercal]\\
 &= \mathrm{I}(\bm{\theta_0})E_x[(\bm{\hat{\theta}} - \bm{\theta_0})(\bm{\hat{\theta}} - \bm{\theta_0})^\intercal] = \mathrm{I}(\bm{\theta_0}) \Sigma 
\end{split}
\end{equation}
leading to 
\begin{equation}
E_x[\ln g(x|\bm{\theta_0})] \approx E_x[\ln g(x|\bm{\hat{\theta}}(x))] - \frac{1}{2} \mathrm{tr}\left( \mathrm{I}(\bm{\theta_0}) \Sigma \right)
\end{equation}
Using this result in Eq.~(\ref{T}) we get
\begin{equation}
T \approx E_x[\ln g(x|\bm{\hat{\theta}}(x))] - \mathrm{tr}\left( \mathrm{I}(\bm{\theta_0}) \Sigma \right)
\end{equation}
Maximizing this expression can be used as a criterion for model selection when we have many and large samples to average over. 

If however we have only one sample at our disposal, we can assume that an estimator of $T$, $\hat{T}$, will be of the same form as $T$, without the expected value and with an estimator for the trace
\begin{equation}
\hat{T} \approx \ln g(x|\bm{\hat{\theta}}) - \hat{\mathrm{tr}}\left(\mathrm{I}(\bm{\theta_0}) \Sigma\right)
\end{equation}
Instead of maximizing $T$, it is a convention to minimize the quantity 
\begin{equation}
-2\hat{T} \approx -2\ln g(x|\bm{\hat{\theta}}) +2  \hat{\mathrm{tr}}\left( \mathrm{I}(\bm{\theta_0})\Sigma\right)
\end{equation}
It can be shown \cite{BA} that if $g(x|\bm{\theta})$ is a good approximation for the true distribution $f(x)$, then 
$\mathrm{I}(\bm{\theta_0}) = \Sigma^{-1}$ and 
$\hat{\mathrm{tr}}\left( \mathrm{I}(\bm{\theta_0})\Sigma\right) = \mathrm{tr}\left( \mathbb{I}_{n \times n} \right) = n$, where $n$ is the number of parameters. Thus, the quantity to be minimized is 
\begin{equation}
\mathrm{AIC} = -2\ln g(x|\bm{\hat{\theta}}) +2 n
\end{equation}
where $\hat{\theta}$ is the maximum likelihood estimate and $n$ the number of parameters of the model.


\subsubsection{Bayesian Information Criterion}
From Bayes' theorem, the posterior probability of model $M_i$ from a set of $r$ candidate models $\{M_1,\dots,M_r\}$, given a set of observations $\bm{x} = \{x_1,\dots,x_n\}$ is
\begin{equation}
\label{b1}
P(M_i|\bm{x}) = \frac{g(\bm{x}|M_i)P(M_i)}{\sum\limits_{j=1}^{r}g(\bm{x}|M_j)P(M_j)}
\end{equation}
where $P(M_i)$ is the prior probability of model $M_i$. The likelihood of $M_i$, $g(\bm{x}|M_i)$ is given by
\begin{equation}
\label{b2}
g(\bm{x}|M_i) = \int g(\bm{x}|\bm{\theta}, M_i)\pi(\bm{\theta}|M_i)d\bm{\theta}
\end{equation}
where $g(\bm{x}|\bm{\theta}, M_i)$ is the likelihood of the model parameters $\bm{\theta} = (\theta_1, \theta_2,...,\theta_n)$ and $\pi(\bm{\theta}|M_i)$ the corresponding prior probabilities, and is referred to as \textit{marginal likelihood} since the parameters are marginalized (integrated) out. It is also known as \textit{model evidence} as it expresses the probability of the data $\bm{x}$ being generated from model $M$ whose parameters are sampled from the specified prior distribution.

It is obvious from Eq.~(\ref{b1}) that the model with the largest posterior probability is the one with the largest $g(\bm{x}|M)P(M)$ and assuming that all models have the same prior probabilities $P(M)$ it is the model with the largest $g(\bm{x}|M)$. The goal is, therefore, to maximize the quantity $g(\bm{x}|M)$ as it hints us to the most probable model from the set of $r$ candidate models $\{M_1,\dots,M_r\}$.

The integral in Eq.~(\ref{b2}) can be evaluated using the Laplace approximation \cite{Bishop, MK}, whereby the logarithm of $g(\bm{x}|\bm{\theta}, M_i)$ in the integrand is Taylor-expanded to second order around its maximum $\bm{\hat{\theta}}$, leading to a Gaussian integral. 
The expansion of the log-likelihood ($M_i$ is omitted for simplicity) yields 
\begin{equation}
\label{b3}
\ln g(\bm{x}|\bm{\theta}) \approx \ln g(\bm{x}|\bm{\hat{\theta}}) - \frac{1}{2} ( \bm{\theta} - \bm{\hat{\theta}} )^\intercal \mathrm{H}(\bm{\theta} - \bm{\hat{\theta}})
\end{equation}
since the gradient vanishes at $\bm{\theta}=\bm{\hat{\theta}}$,
 with $\mathrm{H} = -\nabla \nabla \ln g(\bm{x}|\bm{\hat{\theta}}) |_{\bm{\theta}=\bm{\hat{\theta}}}$ the $n \times n$ Hessian matrix of the log-likelihood. After exponentiation, the likelihood function becomes
\begin{equation}
\label{b4}
g(\bm{x}|\bm{\theta}) \approx  g(\bm{x}|\bm{\hat{\theta}})\exp\left\lbrace-\frac{1}{2} ( \bm{\theta} - \bm{\hat{\theta}} )^\intercal \mathrm{H}(\bm{\theta} - \bm{\hat{\theta}})\right\rbrace
\end{equation}
Assuming a uniform prior, $\pi(\bm{\theta})$ can be considered constant and $\pi(\bm{\theta})\approx \pi(\bm{\hat{\theta}})$, 
so that the right hand side of Eq.~(\ref{b2}) can be calculated as  an $n$-dimensional Gaussian integral resulting in 
\begin{equation}
\label{b5}
g(\bm{x}|M) \approx g(\bm{x}|\bm{\hat{\theta}})\pi(\bm{\hat{\theta}}) \frac{(2\pi)^{n/2}}{\lvert \mathrm{H} \rvert^{1/2}}
\end{equation}
where $\lvert \mathrm{H} \rvert$ is the determinant of H. Since the likelihood of a sample is the product of the likelihoods of each observation, the log-likelihood that appears in the elements of the matrix H is the sum of log-likelihoods of the observations and the elements of H become
\begin{equation}
\label{b6}
\begin{split}
\mathrm{H}_{kl} &= -\left. \frac{\partial^2 \ln g(\bm{x}|\bm{\theta})}{\partial \theta_k \partial \theta_l}\right \rvert_{\bm{\theta} = \bm{\hat{\theta}}} = -\left.\frac{\partial^2 \ln \prod_{j=1}^{N} g(x_j|\bm{\theta})}{\partial \theta_k \partial \theta_l}\right \rvert_{\theta = \hat{\theta}}\\
&= -\left.\frac{\partial^2 \sum_{j=1}^N \ln g(x_j|\bm{\theta})}{\partial \theta_k \partial \theta_l}\right \rvert_{\theta = \bm{\hat{\theta}}}\\
&= -N\left.\frac{\partial^2 E[ \ln g(x_j|\bm{\theta})]}{\partial \theta_k \partial \theta_l}\right \rvert_{\bm{\theta} = \bm{\hat{\theta}}} = N \mathrm{I}(\bm{\hat{\theta}})_{kl}
\end{split}  
\end{equation}
where N is the sample size and $\mathrm{I}(\bm{\hat{\theta}})$ the Fisher information matrix. Thus, the determinant of H equals $|\mathrm{H}| = N^{n}|\mathrm{I}|$, where n is the dimensionality of the parameter space. Finally, if we take again the logarithm of Eq.~(\ref{b5}) we get
\begin{equation}
\label{b7}
\ln g(\bm{x}|M) \approx \ln g(\bm{x}|\bm{\hat{\theta}}) + \ln \pi(\bm{\hat{\theta}}) + \frac{n}{2} \ln 2\pi - \frac{1}{2} n \ln N - \frac{1}{2} \ln |\mathrm{I}|
\end{equation}
and keeping only terms that vary at least linearly with sample size N, we end up with
\begin{equation}
\label{b8}
\ln g(\bm{x}|M) \approx \ln g(\bm{x}|\bm{\hat{\theta}}) - \frac{n}{2} \ln N.
\end{equation} 
Let us recall that $g(\bm{x}|M)$ is the quantity we sought to maximize and that $\bm{\hat{\theta}}$ is the maximum likelihood estimate. As a matter of convention we define
\begin{equation}
\label{b9}
\mathrm{BIC} = - 2 \ln g(\bm{x}|M) \approx \ln g(\bm{x}|\bm{\hat{\theta}}) + n \ln N
\end{equation} 
as the quantity that, when minimized, will yield the most probable model given the data.


\end{document}